\documentclass[11pt]{article}
\usepackage{verbatim}

\newenvironment{fullv}{}{}
\newcommand{\cconfv}[1]{}
\newcommand{\cfullv}[1]{#1}
\newcommand{\Section}[1]{\section{#1}}

\usepackage{amsmath,amsthm,amsfonts}

\renewcommand{\Re}{\mathbb{R}}

 \theoremstyle{plain}
 \newtheorem{theorem}{Theorem}
 \newtheorem{lemma}{Lemma}
 \newtheorem{corollary}[lemma]{Corollary}
 
 \newtheorem{proposition}[lemma]{Proposition}

 \theoremstyle{definition}
 \newtheorem{definition}{Definition}
 
 \newtheorem{observation}[lemma]{Observation}

 \DeclareMathOperator{\dist}{d}
 \newcommand{\diam}{\Delta}
 \DeclareMathOperator{\cost}{cost}
 \DeclareMathOperator{\lca}{lca}

\newcommand{\E}{\mathrm{E}}
\newcommand{\A}{\mathcal{A}}
\newcommand{\e}{\varepsilon}
\newcommand{\ie}{\emph{i.e.}}
\newcommand{\etal}{\emph{et. al.}}

\newcommand{\opt}{\ensuremath{\mathrm{OPT}}}

\newcommand{\be}{\begin{enumerate}}
\newcommand{\ee}{\end{enumerate}}
\newcommand{\bi}{\begin{itemize}}
\newcommand{\ei}{\end{itemize}}

\begin{document}

\title{Ramsey-type Theorems for Metric Spaces with Applications
to Online Problems\thanks{A preliminary version, entitled
``\emph{A Ramsey-type Theorem for Metric Spaces and its
Applications for Metrical Task Systems and Related Problems}",
appeared in Proceedings of the 42nd annual Symposium on
Foundations of Computer Science, 2001.}}

\author{Yair Bartal%
\thanks{Supported in part by a grant from the Israeli Science Foundation (195/02).}
\\ Hebrew University\\ Jerusalem, Israel\\
{yair@cs.huji.ac.il}
 \and B\'{e}la Bollob\'{a}s\\ University of Memphis\\ Memphis, TN 38152\\
{bollobas@msci.memphis.edu}
 \and Manor Mendel\thanks{Work mostly done while the author was a
Ph.D. student in Tel-Aviv University, under
 the supervision of Prof. A. Fiat. Author's current address: Department of
Computer Science, University of Illinois, Urbana, IL 61801, USA.
mendelma@uiuc.edu. Supported in part by a grant from the Israeli
Science Foundation (195/02).}
 \\ Hebrew University\\ Jerusalem, Israel\\  {mendelma@yahoo.com}}

\date{\today}

\maketitle

\thispagestyle{empty}

\begin{abstract}
A nearly logarithmic lower bound on the randomized competitive ratio for the
\emph{metrical task systems} problem is presented. This
implies a similar lower bound for the extensively studied $K$-server
problem. The proof is based on Ramsey-type theorems for metric spaces, that
state that every metric space contains a large subspace which is
approximately a \emph{hierarchically well-separated tree} (and
in particular an ultrametric). These Ramsey-type theorems may be of
independent interest.
\end{abstract}

\Section{Introduction}

This paper deals with the analysis of the performance of
randomized online algorithms in the context of two fundamental
online problems --- metrical task systems and the $K$-server
problem.

A {\sl metrical task system} (MTS), introduced by Borodin, Linial,
and Saks \cite{BLS92}, is a system that may be in one of a set of
$n$ internal states. The aim of the system is to perform a given
sequence of \emph{tasks}. The performance of each task has a
certain cost that depends on the task and the state of the system.
The system may switch states; the cost of such a switch is the
distance between the states in a metric space defined on the set
of states. After a switch, the cost of the service is the one
associated with the new state.

In the {\sl $K$-server problem}, defined by Manasse, McGeoch, and
Sleator~\cite{MMS90}, $K$ mobile servers reside in points of a
given metric space. A sequence of requests for points in the space
is presented to the servers. To satisfy a request, one of the $K$
servers must be moved to the point associated with the request.
The cost of an algorithm for serving a sequence of requests is the
total distance travelled by the servers.

An online algorithm receives requests one by one and must serve
them immediately without knowledge of future requests. A
randomized online algorithm is called $r$-\emph{competitive} if on
every sequence its expected cost is at most $r$ times the optimal
offline cost plus an optional constant additive term.

The MTS and $K$-server problems have been studied extensively with the aim of
determining the best competitive ratio of online algorithms.
Borodin et al. \cite{BLS92} have shown
that the deterministic competitive ratio for MTS on an $n$ point metric
space is exactly $2n-1$. Manasse et al. \cite{MMS90} proved a lower bound of
$K$ on the competitive ratio of deterministic $K$-server algorithms. The
best upper bound for arbitrary metric spaces and any $K$ is currently $2K-1$
\cite{KP95}.

The randomized competitive ratio for these problems is not as well
understood. For the uniform metric space, where all distances are
equal, the randomized competitive ratio is known to within a
constant factor, and is $\Theta(\log n)$
\cfullv{\cite{BLS92,IS98}}\cconfv{\cite{BLS92}} for MTS and
$\Theta(\log K)$ \cfullv{\cite{FKLMSY91,MS91,ACN00}}
\cconfv{\cite{FKLMSY91}} for the $K$-server problem. In fact, it
has been conjectured that, in any metric space, the randomized
competitive ratio is $\Theta(\log n)$ for MTS and $\Theta(\log K)$
for the $K$-server problem. Previous lower bounds were
$\Omega(\log\log n)$ \cite{KRR94}, and $\Omega(\sqrt{\log
n/\log\log n})$ \cite{BKRS00} for MTS and similar lower bounds for
the $K$-server problem in metric spaces with more than $K$ points.
The upper bound for MTS was improved in a sequence of papers
\cite{Bar96,BBBT97,Bar98,FM00,BM03,FRT03}, and is currently
$O(\log^2 n\log\log n)$. The upper bound for MTS implies a similar
bound for the $K$-server problem on $K+c$ points, when $c$ is a
constant. However, no ``general" randomized upper bound for the
$K$-server problem better than $2K-1$ \cite{KP95} is currently
known. Seiden \cite{Sei01} has a result in this direction, showing
sub-linear bounds for certain spaces with certain number of
servers.

In this paper we give lower bounds on the randomized competitive ratios that
get closer to the conjectured bounds. We prove that, in any $n$-point metric
space, the randomized competitive ratio of the MTS problem is $\Omega(\log
n/\log^2\log n)$.  For the $K$-server problem, we prove that the randomized
competitive ratio is $\Omega(\log K / \log^2\log K)$ for any metric space on
more than $K$ points. Slightly better bounds are obtained for specific
metric spaces such as $\ell$-dimensional meshes. We also prove for any
$\e>0$, a lower bound of $\Omega(\log K)$ for the $K$-server problem in any
$n$-point metric space where $n\geq K^{\log ^\e K}$, improving a lower bound
from \cite{KRR94} of $\Omega(\min\{\log K, \log \log n\})$. We note that the
improved lower bounds for the $K$-server problem also imply  improved lower
bounds for the \textsl{distributed paging problem}, as shown in \cite{ABF93}.
The lower bounds for the $K$-server problem follow from a general reduction
from MTS on a metric space of $K+1$ points to the $K$-server problem in the
same metric space. The rest of the discussion is therefore in terms of the
MTS problem.

In \cite{KRR94,BKRS00,Bar96} it is observed that the randomized competitive
ratio for MTS is conceptually easier to analyze on ``decomposable spaces",
spaces that are composed of subspaces with small diameter compared to that
of the entire space. Bartal~\cite{Bar96} introduced a class of decomposable
spaces he called \emph{hierarchically well-separated trees} (HST). A $k$-HST
is a metric space defined on the leaves of a tree such that, for each level
of the tree, the diameters of the subtrees decrease by a factor of $k$
between the levels. Consider a particular level of an HST. The distances to
all subtrees are approximately the same and thus it is natural to use a
recursive solution for the HST where the problem at a particular level is
essentially on a uniform space.

In order to analyze the competitive ratio for a specific metric
space $M$, it is helpful to consider how close it is to a simpler
metric space $N$ (such as HST). We say that $N$
$\alpha$-approximates $M$ if the distances in $N$ are within a
factor $\alpha$ from those in $M$. Clearly, if there is a
$r$-competitive algorithm for $N$ then there is $\alpha
r$-competitive algorithm for $M$. This notion can be generalized
to a probabilistic metric approximation \cite{Bar96} by
considering a set of approximating metric spaces that dominate the
original metric space and bounding the expectation of the
distances. Any metric space on $n$ points can be $O(\log
n)$-\emph{probabilistically approximated} by HSTs
\cite{Bar96,Bar98,FRT03}, thus reducing the problem of devising
algorithm for MTS on any metric space to devising an algorithm for
HSTs only \cite{BBBT97,FM00}. HSTs and their probabilistic
approximation of metric spaces have found many other applications
in online and approximation algorithms, for example
{\cite{Bar96,AA97,KT99}}.
See \cite[Sect.~2.4 and 5]{Ind01} for a survey on this topic.

The first step toward obtaining a lower bound for arbitrary metric spaces is
showing that a lower bound for HSTs implies a lower bound for arbitrary
metric spaces.
Probabilistic approximations are not useful for this purpose. One
of the reasons for this is that the approximation bound is at
least logarithmic, and therefore a logarithmic lower bound for
HSTs would not imply any non trivial lower bound for arbitrary
metrics. What makes the reduction in this paper possible is the
observation that a lower bound for a subspace implies a lower
bound for the entire space. Therefore, in order to get a lower
bound for a specific metric space $M$, we need to find a large
subspace which is a good approximation of an HST. Such theorems
are called \textsl{Ramsey-type theorems for metric
spaces}~\cite{KRR94}. The main Ramsey-type theorem in this paper
states that in any metric space on $n$ points there exists a
subspace of size $n^{\Omega({\log ^{-1}k})}$ points that
$O(\log\log n)$-approximates a $k$-HST. In fact, we further show
that the approximated $k$-HST can have the additional property
that any internal vertex of the underlying tree of the HST, either
has only two children or all the children's sub-trees are of
almost equal size (in terms of the number of leaves). It is worth
noting that HSTs are ultrametrics and thus embed isometrically in
$\ell_2$. Therefore, our Ramsey-type theorems give subspaces in
Euclidean space.
 Previously, Bourgain {\etal}~\cite{BFM86}, Karloff
{\etal}~\cite{KRR94} and Blum {\etal}~\cite{BKRS00} proved other
Ramsey-type theorems, showing the existence of special types of
HSTs on significantly smaller subspaces. In
Section~\ref{sec:additional Ramsey} we elaborate on these results
and relate our constructions to their constructions.
Subsequent work is discussed in
Section~\ref{sec:subsequent}. Different Ramsey-type problems for
metric spaces appear in~\cite{Mat92}.

The lower bound for HST spaces follows a general framework
originated in \cite{BKRS00} and explicitly formulated in
\cite{Sei99,BBBT97}: The recursive structure of the HST is
modelled via  the {\sl unfair metrical task system} (UMTS) problem
\cite{Sei99,BBBT97} on a uniform metric space. This concept is
presented in greater details in Section~\ref{sec:overview}. For
readers already familiar with the concept, the rest of this
paragraph provides a brief summary of our results for this model.
In a UMTS problem, every point $v_i$ of the metric space is
associated with a \emph{cost ratio} $r_i$ which multiplies the
online costs for processing tasks in that point. Offline costs
remain as before. The cost ratio $r_i$ roughly corresponds to the
competitive ratio of the online algorithm in a subspace of the
HST. We prove a lower bound for the randomized competitive ratio
of a UMTS on the uniform metric space \emph{for the entire range
of cost ratios} $(r_i)_i$, $r_i\geq 1$. This lower bound is tight
assuming the conjectured $\Theta(\log n)$ competitive ratio for
MTS. Previously, tight lower bounds (in the above sense) were only
known for two point spaces \cite{BKRS00,Sei99} and fair
MTS\cfullv{, where $r_1=r_2=\cdots=r_n=1$}~\cite{BLS92}.
\emph{Upper bounds} for UMTS problems were given for two point
spaces \cite{BKRS00,Sei99, BBBT97} and when all the cost ratios
are equal $r_1=r_2=\cdots =r_n$ \cite{BBBT97}. Our lower bound
matches these upper bounds in these cases.

By making use of the lower bounds for UMTSs on uniform metric
spaces, we compose lower bounds to obtain a lower bound of
$\Omega(\log n)$ for HSTs. Our main lower bound result follows
from the lower bound on HST and the Ramsey-type theorem.

\subsection{Subsequent Work}\label{sec:subsequent}

Subsequent to this paper, metric Ramsey problems have been further
studied in a sequence of papers
\cite{BLMN-phenomena,BLMN-dichotomy,BLMN-lowdist,BLMN-frechet}.
The main theorem in \cite{BLMN-phenomena} states that any $n$
point metric space contains a subspace of size $n^{1-\frac{c\log
\alpha}{\alpha}}$ which $\alpha$-approximates a $1$-HST for
$\alpha>2$ and an appropriate value $c>0$. Since a $1$-HST is
equivalent to an ultrametric which isometrically embeds in
$\ell_2$ this theorem gives nearly tight Ramsey-type theorem for
embedding metrics spaces in Euclidean space.
 The proof of the
theorem uses techniques developed in this paper, but is more
involved and requires new ingredients as well.

 It follows from \cite{BLMN-phenomena} that the main Ramsey theorem in
this paper (Theorem~\ref{thm:khst-subspace}) can be improved to
the following: there exists $c>0$ such that any $n$ point metric
space contains a subset of size $n^{c/\log(2k)}$ which
3-approximates a $k$-HST. Together with the lower bounds for
$k$-HSTs in the current paper (Theorem~\ref{thm:hst-lb}), the
lower bound on the randomized competitive ratio for the MTS
problem on $n$-point metric space (Theorem~\ref{thm:main}) is
improved to $\Omega(\log n / \log\log n)$, and the lower bound for
the $K$-server problem is improved to $\Omega(\log K/ \log\log
K)$.  Also, the results of Section~\ref{sec:additional Ramsey} are
complemented in \cite{BLMN-dichotomy}, where tight bounds on these
metric Ramsey problems are given.

Also related is work on Multi-embeddings \cite{BM03} which studies
a concept that can be viewed as dual to the Ramsey problem. In a
multi-embedding a metric space is embedded in a larger metric
space where points embed into multiple points. This paper as well
uses techniques very similar to the ones developed here. It is
also shown how this concept can be used to obtain upper bounds for
the MTS problem.

\paragraph{Outline of the paper.}
In Section~\ref{sec:overview} the problems and the main concepts are
formally defined along with an outline of the proof of the lower bound.
Section~\ref{sec:ramsey} is devoted to present our main Ramsey-type theorem
for metric spaces. In Section~\ref{sec:unfair} we prove a lower bound for
UMTSs on a uniform metric spaces, and use it in
Section~\ref{sec:hst-lb} to deduce a lower bound for HSTs. In
Section~\ref{sec:servers} we apply these lower bounds to the $K$-server
problem. In Section~\ref{sec:additional Ramsey} we discuss additional
Ramsey-type theorems and tight examples. We also relate our work to previous
known constructions. Finally, in Section~\ref{sec:open}, we present a number
of open problems that arise from the paper.

\Section{Overview and Definitions } \label{sec:overview}

In this section we outline the proof of the lower bounds for the metrical
task systems problem on arbitrary metric spaces. We start with defining the
MTS problem.

\begin{definition}
A metric space $M=(V,d)$ consists of a set of points $V$ and a metric
distance function $d:V\times V \rightarrow \Re^+$ such that $d$ is
symmetric, satisfies the triangle inequality and  $d(u,v) = 0$ if and only if $u=v$.

For $\alpha>0$, we denote by $\alpha M$ the metric space obtained
from $M$ by scaling the distances in $M$ by a factor $\alpha$.
\end{definition}

\begin{definition}
A \emph{metrical task system} (MTS) \cite{BLS92} is a problem
defined on a metric space $M=(V,\dist _M)$ that consists of
$|V|=b$ points, $v_1,\ldots,v_b$.  The associated online problem
is defined as follows. Points in the metric space represent
internal states of an online algorithm $A$. At each step, the
algorithm $A$ occupies a point $v_i\in M$. Given a task, the
algorithm may move from $v_i$ to a point $v_j$ in order to
minimize costs. A task is a vector $(c_1,c_2, \ldots, c_b) \in
\bigl (\mathbb{R}^+ \cup \{\infty\}\bigr )^b$, and the cost for
algorithm $A$ associated with servicing the task is
$\dist_M(v_i,v_j) + c_j$. The \emph{cost} for $A$ associated with
servicing a sequence of tasks $\sigma$, denoted by
$\cost_A(\sigma)$, is the sum of costs for servicing the
individual tasks of the sequence consecutively. and is denoted by
$\cost_A(\sigma)$. An online algorithm makes its decisions based
only upon the tasks seen so far.
\end{definition}

We define $\cost_{\text{OPT}}(\sigma)$ to be the minimum cost, for any
off-line algorithm, to start at the initial state and process $\sigma$. A
randomized online algorithm $A$ for an MTS is an online algorithm that
decides upon the next state using a random process. The expected cost of a
randomized algorithm $A$ on a sequence $\sigma$ is denoted by
$\E[\cost_A(\sigma)]$.

\begin{definition} \cfullv{\cite{SleTar85a,KMRS88,BBKTW94}} \label{def:competitive}
A randomized online algorithm is called $r$-\emph{competitive
against an oblivious adversary} if there exists a constant $c$
such that for every task sequence $\sigma$, $\E[\cost_A(\sigma)]
\leq r \cdot \cost_{\opt}(\sigma) + c$.
\end{definition}

The main result of this paper is the following theorem.

\begin{theorem} \label{thm:main}
Given a metric space $M$ on $n$ points,
the competitive ratio (against oblivious adversaries) of any randomized online
algorithm for the MTS defined on $M$, is at least
$\Omega(\log n/ \log^2 \log n)$.
\end{theorem}

We first observe the fact that a lower bound for a sub-space of $M$ implies
a lower bound for $M$. Therefore if we have a class of metric spaces
$\mathcal{S}$ for which we have a lower bound we can get a lower bound for a
metric space $M$ if it contains a metric space, $M' \in \mathcal{S}$ as a
subspace. This may also be done if the subspace approximates the metric
space $M'$.

\begin{definition} \label{def:metric-approx}
A metric space $M$ over $V$ $\alpha$-approximates a metric space $M'$ over
$V$ if for all $u,v\in V$,
  \(
 d_{M'}(u,v) \leq d_M(u,v) \leq \alpha d_{M'}(u,v) .
  \)
\end{definition}

Note that Definition~\ref{def:metric-approx} is essentially
symmetric in a sense that if $M$ $\alpha$-approximates $M'$, then
$M'$ $\alpha$-approximates $\alpha^{-1} M$.

\begin{proposition} \label{prop:metric_approx_lb}
Given a metric space $M$ on $V$ that
  $\alpha$-approximates a metric space $M'$ on $V$,
a lower bound of $r'$ for the MTS on $M'$ implies a lower bound of\/ $r'/
\alpha$ for the MTS on $M$.
\end{proposition}
\begin{fullv}
\begin{proof}
Assume there exists an $r$-competitive algorithm $A$ for $M$. Let $A'$ be
the algorithm that simulates $A$ on $M'$. Let $B'$ be an optimal algorithm
for $M'$ and let $B$ be its simulation in $M$. Then
\begin{equation*}
\E[\cost_{A'}(\sigma)] \leq \E[\cost_{A}(\sigma)]
    \leq r\, \cost_{B}(\sigma) + c
    \leq \alpha r\, \cost_{B'}(\sigma) +\alpha c.
\end{equation*}
Therefore $A'$ is $\alpha\, r$ competitive.
\end{proof}
\end{fullv}

Next, we define the class of metric spaces for which we will construct lower
bounds for the MTS problem. Following Bartal~\cite{Bar96}\footnote{The
definition given here for $k$-HST differs slightly from the original
definition in \cite{Bar96}.
For $k >1$ the metric spaces given by these
two definitions approximate each other to within a factor of $k/(k-1)$.}, we
define the following class of metric spaces.

\begin{definition}
For $k\geq 1$, a $k$-\emph{hierarchically well-separated tree} ($k$-HST) is
a metric space defined on the leaves of a rooted tree $T$. To each vertex
$u\in T$ there is associated a label $\Delta(u) \ge 0$ such that
$\Delta(u)=0$ if and only if $u$ is a leaf of $T$. The labels are such that if a vertex
$u$ is a child of a vertex $v$ then $\Delta(u)\leq \Delta(v)/k$. The
distance between two leaves $x,y\in T$ is defined as $\Delta(\lca(x,y))$,
where $\lca(x,y)$ is the least common ancestor of $x$ and $y$ in $T$.
Clearly, this function is a metric on the set of vertices. We call a vertex
with exactly one child, a \emph{degenerate vertex}. For a non-degenerate vertex
$u$, $\Delta(u)$ is the diameter of the sub-space induced on the subtree
rooted by $u$. Any $k$-HST can be transformed into a $k$-HST without
degenerate vertices and with the same metric.
\end{definition}

Any $k$-HST is also a $1$-HST.
We use the term \emph{HST} to denote any $1$-HST.
An HST is usually referred to as \emph{ultrametric}, but note
that for $k>1$, a $k$-HST is a stronger notion.

In Section~\ref{sec:ramsey} we prove a generalized form of the
following Ramsey-type theorem for metric spaces.
\begin{theorem} \label{thm:khst-subspace}
Given a metric space $M=(V,d)$ on $|V|=n$ points and a number $k \ge 2$,
there
exists a subset $S\subseteq V$ such that $|S|\geq n ^{\Omega(1/\log k)}$ and
the metric space $(S,d)$ $O(\log\log n)$-approximates a $k$-HST.
\end{theorem}

It follows that it suffices to give lower bounds for the MTS
problem on HST metric spaces. A natural method for doing that is
to recursively combine lower bounds for subspaces of the HST into
a lower bound for the entire metric space. Consider an internal
vertex $u$ at some level of the HST. Let $v_1, v_2, \ldots, v_b$
be its children and assume we have lower bounds of $r_1, r_2,
\ldots, r_b$ on the competitive ratio for the subspaces rooted at
the $v_i$s. We would like to combine the lower bounds for these
subspaces into a lower bound of $r$ for the subspace rooted at
$u$. Recall that the distances between points in the subspaces
associated with different $v_i$s are equal to $\diam(u)$. We would
like to think of such a subspace rooted at $v_i$ as being replaced
by a single point and the subspace rooted at $u$ being a uniform
metric space.  Given a task of cost $\delta$ at the point
associated with the subspace rooted at $v_i$ the cost charged to
the online algorithm is at least $r_i\delta$. Informally speaking,
given a lower bound for this metrical task system with unfair
costs on a uniform metric space we can obtain a lower bound for
the subspace rooted at $u$. This serves as a motivation for the
following definition.

\begin{definition} \cite{BKRS00,Sei99,BBBT97}
An \emph{unfair metrical task system} (UMTS) $U=(M;r_1,\ldots,r_b;s)$
consists of a metric space $M$ on $b$ points, $v_1,\ldots,v_b$, with a
metric $\dist_M$, a sequence of \emph{cost ratios} $r_1,r_2, \ldots,r_b\in
\Re^+$, and a \emph{distance ratio} $s\in \Re^+$. For $s=1$, we omit the
parameter $s$ from the notation.

The UMTS problem differs from the regular MTS problem in that the
cost of the online algorithm for servicing a task $(c_1,c_2,
\ldots, c_b)$ by switching from $v_i$ to $v_j$ is $s\cdot
\dist_M(v_i,v_j) + r_j c_j$, whereas the offline cost remains as
before.
\end{definition}

\begin{observation} \label{obs:w/o-dr}
It is sufficient to analyze UMTSs with distance ratio equals one
since a UMTS $U=(M;r_1,\ldots,r_b;s)$ has a competitive ratio $r$
if and only if $U'=(M;r_1s^{-1},\ldots,r_bs^{-1};1)$ has a
competitive ratio $rs^{-1}$. This is so since the adversary costs
in $U$ and $U'$ are the same, whereas the online costs in $U$ are
$s$ times larger than in $U'$.
\end{observation}

Our goal is to obtain lower bounds for the UMTS problem on a
uniform metric space (where all distances between different points
are equal). Consider attempting to prove an $\Omega(\log n)$ lower
bound for fair MTS problem on HST metric. If our abstraction is
correct, it is reasonable to expect that for UMTS
$U=(\mathcal{U}_b^\Delta;r_1,\ldots,r_b)$ (where
$\mathcal{U}_b^\Delta$ is the uniform metric space on $b$ points
with distance $\Delta$), if $r_i\geq c\log n_i$ then there is a
lower bound of $r$ for $U$ such that $r\geq c\log (\sum_i n_i)$.
Indeed we prove such a claim in Section~\ref{sec:unfair}
(Lemma~\ref{lem:adv-composed}).
In Section~\ref{sec:hst-lb} we combine the lower bounds for the uniform UMTS
and obtain an
$\Omega(\log n)$ lower bound on a $k$-HST along the lines outlined above. In
order to avoid interference between the levels this applies only for
$k=\Omega(\log^2 n)$. We prove

\begin{theorem} \label{thm:hst-lb}
Given an $\Omega(\log^2 n)$-HST $M$ on $n$ points,
the competitive ratio (against oblivious adversaries) of any
randomized online algorithm for the MTS defined on $M$, is at least
$\Omega(\log n)$.
\end{theorem}

Theorem~\ref{thm:hst-lb}, Theorem~\ref{thm:khst-subspace} and
Proposition~\ref{prop:metric_approx_lb} imply a lower bound of $\Omega(\log
n/\log^2\log n)$ for any metric space, which concludes Theorem \ref{thm:main}.

\Section{Ramsey-type Theorems for Metric Spaces } \label{sec:ramsey}

\begin{lemma} \label{lemma:HPM-subspace}
Given a metric space $M=(V,d)$ on $|V|=n$ points, and  $\beta > 1$, there
exists a subset $S\subseteq V$, such that $|S|\geq n ^{1/\beta}$ and
$(S,d)$ $O(\log_\beta \log n)$-approximates a $1$-HST.
\end{lemma}
\begin{proof}
We may assume that $\beta \le \log n$ and $n>2$ (otherwise the claim is
trivial). Let $\Delta$ be the diameter of $M$, and let $t= \lceil \log_\beta
\log n +1 \rceil$. Choose an endpoint of the diameter $x\in M$. Define a
series of sets $A_i=\{y\in M |\ d(x,y)\leq \Delta (i/(2t+1)\}$, for
$i=0,1,2,\ldots,2t+1$, and ``shells" $S_0=\{x\}$, $S_i=A_i\setminus
A_{i-1}$. Choose $S_i$, \(1\leq i \leq 2t\), and delete it. Denote by
$B=A_{i-1}$ and $C=V\setminus A_{i}$. The root of the $1$-HST is associated
with label $\Delta/(2t+1)$, the two sub-trees are built recursively by
applying the same procedure on $(B,d|_{B})$ and $(C,d|_C)$. Let $S$ be the
resulting set of points that are left at the end of the recursive process.
Since distances in the $1$-HST are at most $1/(2t+1)$ smaller than those in
$M$ we get that the subspace $(S,d)$ indeed $2t+1=O(\log_ \beta \log n
)$-approximates the resulting $1$-HST. We are left to show how to choose $i$
such that $|S|\geq n^{1/\beta}$.

Let $\e_i= A_i/n$. Note that $n^{-1}\leq \e_0 \leq \e_{2t} \leq 1-
n^{-1}$. Without loss of generality we may assume that $\e_t\leq
1/2$, since otherwise we may consider the sequence
$A'_i=V\setminus A_{2t-i}$ and $\e'_i= 1- \e_{2t-i}$. After
deleting $S_i$ we are left with two subspaces, $|B|=\e_{i-1}n$,
and $|C|=(1-\e_i)n$. Inductively, assume that the recursive
selection leaves at least $(\e_{i-1}n)^{1/\beta}$ points in $B$
and at least $((1-\e_{i})n)^{1/\beta}$ points in $C$. So  \(
|S|\geq (\e_{i-1}^{1/\beta} + (1-\e_i)^{1/\beta}) n^{1/\beta}\)
points. To finish the proof it is enough to show the existence of
$i_0$ for which $\e_{i_0-1}^{1/\beta} + (1-\e_{i_0})^{1/\beta}
\geq 1$. If exists $0\leq i_0 <t $ for which $\e_{i_0-1} \geq
\e_{i_0}^\beta$ then \( \e_{i_0-1}^{1/\beta}+
(1-\e_{i_0})^{1/\beta} \geq \e_{i_0} + (1- \e_{i_0})=1 \) and we
are done. Otherwise, we have that $\e_{i-1} < \e_{i}^\beta$ for
all $0\leq i <t$, and since $\e_t\leq 1/2$, we conclude by
induction on $i$ that $\e_i \leq (1/2)^{\beta^ {(t-i)}}$. But then
\[\e_0 \leq (\tfrac{1}{2})^{\beta^ t}<
   (\tfrac{1}{2})^{\beta^ {\log_\beta \log n}}=\frac{1}{n} , \]
which contradicts $\e_0\geq 1/n$.
\end{proof}

We also need the following lemma from \cite{Bar98}.
\begin{lemma} \label{lem:1-hst}
For any $\ell>1$, any $1$-HST  $\ell$-approximates some
$\ell$-HST.\footnote{In \cite{Bar98}, $1$-HST is referred as ``hierarchical
partition metric".}
\end{lemma}
\begin{proof}[Proof sketch.]
Let $T$ be a $1$-HST. We construct a $\ell$-HST by incrementally changing
$T$ as follows. Scan the vertices of $T$ in top-down fashion. For any
non-root vertex $v$, and its father $u$, if $\Delta(v)\geq \Delta(u)/\ell$
then delete $v$ and connects $v$'s children directly to $u$. The resulting
tree is clearly an $\ell$-HST and a $\ell$-approximation of $T$.
\end{proof}

Next, we show how to prune an $\ell$-HST on $n$ leaves, to get a subtree
which is a $k$-HST with $n^{1/ \lceil \log _\ell k \rceil}$ leaves. This
follows from the following combinatorial lemma for arbitrary rooted trees.
Recall that a vertex in a rooted tree is called non-degenerate if the number
of its children is not one.

\begin{definition}
A rooted tree is  $h$-\emph{sparse} if the number of edges along
the path  between any two non-degenerate vertices is at least $h$.
\end{definition}

\begin{lemma} \label{lem:sparse-tree}
 Given a rooted tree $T$ on $n$ leaves, there exists a subtree $T'$ with at
least $n^{1/h}$ leaves that is $h$-sparse.
\end{lemma}
\begin{proof}
For a tree $T$ and $i\in\{0,1,\ldots,h-1\}$ let $f_i(T)$ be the
maximum number of leaves in $h$-sparse subtree of $T$ for which
any vertex of depth less than $i$ has out degree at most one.
Clearly $f_0(T)=\max_i f_i(T)$.

We prove by induction on the height of $T$ that $\prod_{i=0}^{h-1}
f_i(T) \geq n$, and thus $f_0(T)\geq n^{1/h}$. The base of the
induction is a tree $T$ of height $0$, for which $f_i(T)=1$ for
any $i$, as required. For $T$ with height at least $1$, denote by
$\{T_j\}_j$ the subtrees of $T$ rooted at the children of the root
of $T$. Assume that $T_j$ has $n_j$ leaves, and $n=\sum_j n_j$.
One possible way to obtain an $h$-sparse subtree of $T$ would be
to include the root in the tree and the union of the solutions of
$f_{h-1}(T_j)$. Therefore
\[ f_0(T) \geq \sum_j f_{h-1}(T_j) .\]
Consider the case  $i>0$. Let $v_j$ be a child of the root and let
$T_j$ be the subtree rooted at $v_j$. Let $S_j$ be an $h$-sparse
subtree of $T_j$, with maximum number of leaves, for which any
vertex of depth less than $i-1$ has out degree at most one.
Construct a subtree $S$ by concatenating the edge from the root to
$v_j$ with the subtree $S_j$. This results in an $h$-sparse
subtree of $T$ for which any vertex of depth less than $i$ has out
degree at most one. Hence
\[ f_i(T) = \max_j f_{i-1} (T_j) \qquad \forall i\in \{1,\ldots,h-1\} \]

Thus
\begin{multline*}
\prod_{i =0}^{h-1} f_i(T) \geq \Bigl (\sum_j f_{h-1}(T_j)\Bigr )
\cdot \prod_{i=1}^{h-1} \max_j f_{i-1}(T_j)
 \cconfv{ \\ }
 \geq \sum_j \Bigl (f_{h-1}(T_j) \cdot \prod_{i=1}^{h-1} f_{i-1}(T_j)  \Bigr )
 \\
  = \sum_j \prod_{i=0}^{h-1} f_i(T_j) \geq \sum_j n_j =n
\end{multline*}
The last inequality follows from the induction hypothesis.
\end{proof}

\begin{lemma}\label{lemma:hst-subtree-approx}
Given a $1$-HST $N$ on $n$ points there exists a subspace of $N$ on
$n^{\frac{1}{\lceil \log_\ell k \rceil}}$ points which $\ell$-approximates a
$k$-HST .
\end{lemma}
\begin{proof}
As a first step we construct, using Lemma~\ref{lem:1-hst}, an $\ell$-HST $M$
that is $\ell$ approximated by $N$.

Let $h=\lceil \log_\ell k \rceil$. Let $T$ be the underlying tree
of $M$. Applying Lemma~\ref{lem:sparse-tree} on $T$ we get a
subtree $S$ of $T$ which is $h$-sparse. Let $S'$ be the tree
resulting from coalescing pairs of edges with a common degenerate
vertex in $S$. Consider the metric space $M'$ defined on the
leaves of $S'$ with the associated labels. Clearly, $M'$ is a
subspace of $M$. Consider any internal node $u$ in $S'$ and let
$v$ be a child of $u$ in $S'$. If $v$ is a leaf then $\Delta(v) =
0$. Otherwise both $u$ and $v$ are non-degenerate and therefore
the number of edges  on the path in $T$ between $u$ and $v$ is at
least $h$. This implies that $\Delta(u)/\Delta(v) \ge \ell^h \ge
k$. Thus $M'$ is a $k$-HST.
\end{proof}

\begin{theorem} \label{thm:gen-khst-subspace}
For any metric space $M=(V,d)$ on $|V|=n$ points, any $\beta
>1$, any $k>1$, and any $1< \ell \leq k$
there exists a subset $S\subseteq V$, such that $|S|\geq n
^{\frac{1}{\beta \lceil \log_\ell k \rceil}}$ and $(S,d)$  $O(\ell
\log_\beta \log n)$-approximates a $k$-HST.
\end{theorem}
\begin{proof}
Given a metric space $M$ on $n$ points, from Lemma~\ref{lemma:HPM-subspace},
we get a subspace of $M$ with $n^{1/\beta}$ points that $O(\log_\beta \log
n)$ approximates an $1$-HST $S$. We then apply
Lemma~\ref{lemma:hst-subtree-approx} to obtain a subspace of $S$ on
$n^{\frac{1}{\beta\lceil \log_\ell k \rceil}}$ points which $O(\ell
\log_\beta \log n) $-approximates a $k$-HST.
\end{proof}

Theorem~\ref{thm:khst-subspace} is a corollary of
Theorem~\ref{thm:gen-khst-subspace} when substituting
$\beta=\ell=2$.

As discussed in Section~\ref{sec:subsequent},
Lemma~\ref{lemma:HPM-subspace} has been recently improved in
\cite{BLMN-phenomena}. Lemma~\ref{lemma:hst-subtree-approx} is
tight as shown in Proposition~\ref{prop:tight}. Furthermore we
show in Proposition~\ref{prop:line-inapprox} that in order to get
a Ramsey-type theorem with a constant approximation for HSTs, the
subspace's size must be at most $n^c$ for some constant
$c\in(0,1)$.

For specific metric spaces, better approximations are possible.
Here we consider the $\ell$-dimensional mesh. The result is based on
the Gilbert-Varshamov bound from coding theory (see~\cite[Ch.~17, Thm.
30]{McS77}).

\begin{lemma}[Gilbert-Varshamov bound] \label{lem:GV-bound}
For any $h\in \mathbb{N}$, and $\alpha\in(0,0.5)$, there exists a
binary code $C\subset \{0,1\}^h$ on $h$-bit words such that the
minimum Hamming distance between any two codewords is at least
$\alpha h$, and $|C|\geq 2^{h(1-H_2(\alpha))}$, where $H_2(x)=- (x
\log_2 x + (1-x)\log_2(1-x))$ is the binary entropy.
\end{lemma}

\begin{lemma} \label{lem:mesh}
Given an $h$-dimensional mesh $M=[s]^h=\{0,1,\ldots,s-1\}^h$ with
the $\ell_p$-norm ($p\in[1,\infty]$) on $n=s^h$ points. Then,
there exists a subspace $S\subset [s]^d$ that $12$-approximates a
$9$-HST, and $|S|\geq n ^c$ for  a constant $c=0.08 \log_9 2$.
\end{lemma}
\begin{proof}
We construct an HST $T$ by induction on $s$. For $s=1$, $T$ is simply one
point.

For $s>1$ we construct $T$ as follows. Fix $\alpha=\frac{1}{3}$.
By Lemma~\ref{lem:GV-bound}, there exists an $h$-bit binary code
$C$ with a minimum Hamming distance of $\frac{h}{3}$, and $|C|\geq
2^{h(1-H_2(1/3))}\geq 2^{0.08h}$. For each codeword
$w=(a_1,\ldots, a_h)\in C$, we choose a sub-mesh of size $\lceil
\tfrac{s}{9} \rceil^ h$ with a corner located at $(s-1)w$. More
specifically,
\[ S_w= \bigl(a_1 (s-
\lceil\tfrac{s}{9}\rceil)+\bigl[\lceil\tfrac{s}{9}\rceil\bigr]\bigr)
\times \bigl(a_2 (s-
\lceil\tfrac{s}{9}\rceil)+\bigl[\lceil\tfrac{s}{9}\rceil \bigr ]
\bigr) \times \cdots \times \bigl(a_h (s-
\lceil\tfrac{s}{9}\rceil)+\bigl [\lceil\tfrac{s}{9}\rceil
\bigr]\bigr),\] where for a set of numbers $Y$ and a number $x$,
$x+Y=\{x+y| y\in Y\}$ is the Minkowski sum.

Let $x \in S_w$ and
$y\in S_{w'}$ where $w,w'\in C$ and $w \neq w'$. Obviously
$d(x,y)\leq \sqrt[p]{h}\, (s-1)$, but also, by the triangle
inequality, $d(x,y)$ is at least
\begin{multline*}
d(x,y) \geq (\max_{a\in S_w,\; b\in S_{w'}} d(a,b)) - \text{diam}(S_w)
 - \text{diam}(S_{w'}) \geq \\
  \sqrt[p]{\frac{h}{3}}\,(s-1)- 2
\sqrt[p]{h} (\lceil \frac{s}{9}\rceil -1) \geq
\begin{cases}
\frac{\sqrt[p]{h}}{3}\,(s-1)- 2 \sqrt[p]{h} \frac{s}{9} \geq
\frac{\sqrt[p]{h}\, (s-3)}{9} \geq \frac{\sqrt[p]{h}\, (s-1)}{12}
&
s\geq 9 \\
\frac{\sqrt[p]{h}}{3}\,(s-1) & 2\leq s <9 .
\end{cases}
\end{multline*}
Hence, the distances between points in different sub-spaces are
approximately the same, up-to a factor of $12$. $T$ has a root
labelled with $\frac{\sqrt[p]{h} (s-1)}{12}$. Its children
correspond to the sub-spaces $S_w$ for $w\in C$. For each
sub-space an HST is constructed inductively with $s\leftarrow
\lceil \tfrac{s}{9} \rceil$.

From the construction, $T$ is a $9$-HST, and from the previous
discussion, the distances in $T$ are $12$ approximated by the
original distances in the mesh. $T$ is also a complete and
balanced tree. Its height is at least $\log_9 s$, and the
out-degree of each internal vertex is $|C|$. Hence, the number of
leaves in $T$ is at least $|C|^{\log_9 s} \geq 2^{0.08 h \log_2 s
\log_9 2} = n^c$.
\end{proof}

In Proposition~\ref{prop:line-inapprox} we show the above lemma to be tight.


\Section{Lower Bounds for Uniform UMTS} \label{sec:unfair}

\newcommand{\phases}{\textsc{Phase}}
\newcommand{\D}{\mathcal{D}}
\newcommand{\U}{\mathcal{U}}
\newcommand{\sopt}{\ensuremath{\mathrm{OPT}^0}}

Our goal is to construct a lower bound on HSTs. This is done in the next
section by combining lower bounds for subtrees of the HST, using a lower
bound for a corresponding unfair MTS problem on a uniform metric space. In
this section we formally define the type of the lower bounds we use, and
prove such a lower bound for the uniform metric space.

Our lower bounds are based on Yao's principle (Theorem~\ref{thm:yao}), by
which adversaries produce a distribution over sequences against
deterministic algorithms. However, since the adversaries for (sub)spaces
would be part of a larger adversary, we need to be more careful about their
structure. In particular, since the expected cost of the adversary on the
distribution would serve as the task for UMTS abstracting a higher level
view of the space, and since the lower bounds for UMTS rely crucially on the
tasks being relatively small, we need to maintain upper bounds on the
expected cost of the optimal offline algorithm. We formalize it in the
following definitions.

Given an algorithm $A_U$ for UMTS $U=(M;r_1,\ldots,r_b)$, define
$\cost_{A_U}(\sigma,u_0)$ to be the cost of $A_U$ on the task
sequence $\sigma$ when starting from point $u_0\in M$. Let {\sopt}
be the optimal offline algorithm for servicing a task sequence and
returning to the starting point. An elementary task $(v,\delta)$,
where $v \in M$, is a task that assigns cost $\delta$ to the point
$v$ and $0$ to every other point. Our lower bound argument uses
only elementary tasks.

\begin{definition}
Given a UMTS $U=(M;r_1,\ldots,r_b)$ on a metric space with diameter $\Delta>0$,
define an $(r,\beta)$-\emph{adversary} $\D$ to be a distribution on finite
elementary task sequences for $U$ such that
 \begin{itemize}
\item  \(  \min _{u_0\in M} \E_{\sigma\in \D} [\cost_{\sopt}(\sigma,u_0) ] \le \beta \Delta .\)
\item For any online algorithm $A$,
 \( { \min_{u_0 \in M} \E_{\sigma\in \D}[\cost_A(\sigma,u_0)]} \geq r { \beta \Delta} .
 \)
\end{itemize}
\end{definition}

Yao's Principle (cf. \cfullv{\cite{BLS92,BEY98}}\cconfv{\cite{BLS92}}), as
applied to (unfair) metrical task systems implies the following result.
\begin{theorem} \label{thm:yao}
If there exists an $(r,\beta)$-adversary for a UMTS $U$, then $r$
is a lower bound on the randomized competitive ratio for $U$
against oblivious adversaries.
\end{theorem}
\begin{proof}
The proof is standard and can be found, \emph{e.g.}, in \cite{BEY98}. The
only issue here is to generate a sequence of unbounded cost for the online
algorithm. As we can repeatedly and independently sample from the same
distribution over and over again, we can make the cost of the online
unbounded. Note that the offline costs indeed sum up as required since
{\sopt} always return to the same point.
\end{proof}

Our basic adversaries can only use discrete tasks. We formalize it in the
following definition.

\begin{definition}
Given a UMTS $U=(M;r_1,\ldots,r_b)$ on a metric space with diameter $\Delta$,
an $(r,\beta; \alpha_1,\ldots,\alpha_b)$-\emph{discrete adversary} is an
$(r, \beta)$-adversary that uses only tasks of the form $(v_i,\alpha_i
\Delta)$.
\end{definition}

\begin{observation} \label{obs:scaling}
For $\gamma>0$, denote by $\gamma M$ a metric in which the distances are
scaled by a factor of $\gamma$ compared to $M$. A UMTS
$U=(M;r_1,\ldots,r_b;s)$ and a UMTS $U'=(\gamma M;r_1,\ldots,r_b;s)$ have the
same competitive ratio. Moreover  $(r,\beta)$-adversary and
$(r,\beta;\alpha_1,\ldots,\alpha_b)$-discrete adversary for $U$ are easily
transformed into $(r,\beta)$-adversary and
$(r,\beta;\alpha_1,\ldots,\alpha_b)$-discrete adversary (respectively) for
$U'$ by scaling the tasks by a factor of $\gamma$.
\end{observation}

\begin{lemma} \label{lem:adv-all-cr}
There exist constants\footnote{The constant $\rho$ we achieve is
quite small. We have made no serious attempt to optimize it, and
preferred simplicity whenever possible.} $\lambda_3,\rho
>0$ such that for any UMTS $U=(\mathcal{U}^{\Delta}_b;r_1,\ldots,r_b)$,
$r_1\geq r_2 \geq \cdots \geq r_b \geq 1$ satisfying $r_1\geq \frac{1}{4} \ln
b$, there exists an $(r,\beta; r_1^{-1},r_2^{-1},\ldots,r_b^{-1})$-discrete
adversary, where $\beta \leq \lambda_3  r_1$, and
 \begin{equation} \label{eq:adv-cr}
  r \geq \rho  \ln \Bigl (\sum_{i=1}^b e^{\rho^{-1} r_i} \Bigr ).
 \end{equation}
\end{lemma}

Formula~\eqref{eq:adv-cr} is better understood in the following context. Let
$n_i= e^{\rho^{-1} r_i}$, where $n_i$ should be thought of as a (lower bound)
estimate on the number of points in the subspace that corresponding to
$v_i$, and ``generates" a lower bound on the competitive ratio of
$r_i= \rho \log n_i$. Let $n=\sum_i n_i$,
and thus Formula~\eqref{eq:adv-cr} implies a lower bound of $\rho \ln n$ on
the competitive ratio for the whole space, represented by $U$. This is the
recursive argument we need in order to prove a $\rho \log n$ lower bound.

Without loss of generality (due to Observation~\ref{obs:scaling}),
we may assume that $\Delta=1$. To prove Lemma~\ref{lem:adv-all-cr}
we use the following distribution. Let $m$ be a parameter to be
determined later. A task sequence of length $m$ is generated by
repeatedly and independently picking a random point $v_i$ and
generating an elementary task $(v_i, r_i^{-1})$. The expected cost
of any online algorithm on this distribution is at least $\mu=
\frac{m}{n}$.

We give an upper bound for {\sopt} on this sequence by presenting the
following offline algorithm {\phases}. {\phases} starts at $v_1 \in
\mathcal{U}^{1}_b$. It chooses in hindsight a point $v_i$, moves to $v_i$ at
the beginning of $\sigma$, stay there for the entire duration of $\sigma$,
and at the end returns to $v_1$. The point $v_i$ is chosen so as to minimize
the cost of {\phases}, {\ie}, the local cost on $v_i$ during $\sigma$ plus
zero if $i=1$ and plus two if $i>1$. Denote by $X_i$ the number of tasks
given to point $v_i$. Thus the total local cost for $v_i$ is $X_i/r_i$, and
the expected cost of {\phases} (which is an upper bound on the cost of
{\sopt}) is
 \begin{equation} \label{eq:phase-cost}
 E \Bigl [ \min \bigl \{ \frac{X_1}{r_1}, 2+ \min_{i\geq 2}
\frac{X_i}{r_i} \bigr \} \Bigr ] .
\end{equation}

The analysis of Formula~\eqref{eq:phase-cost} is rather complicated.
Fortunately, in order to prove Inequality~\eqref{eq:adv-cr}, it is
sufficient to establish it in only two cases: when $b=2$ and when
$r_1=\cdots=r_b$. This is due to the following proposition.

\begin{proposition}\label{prop:binary/balanced}
Given a non-increasing sequence of positive real
numbers $(n_i)_{i\geq 1}$. Denote by $n=\sum_i n_i$ and assume
$n<\infty$. Then either $\sqrt{n_1} + \sqrt{n_2} \geq \sqrt{n}$, or there
exists $\ell\geq 3$ such that $\ell \cdot \sqrt{n_\ell} > \sqrt{n}$.
\end{proposition}
\begin{proof}
We first normalize by setting $x_i=n_i/n$. Thus, $\sum_{i} x_i
=1$, and we need to prove that either $\sqrt{x_1} + \sqrt{x_2} \geq 1$ or
there exists $\ell\geq 3$ such that $\ell \sqrt{x_\ell} > 1$.

Assume that the second condition does not holds, {\ie} $\forall
\ell\geq 3$, $x_\ell \leq \ell^{-2}$. we will prove that
$\sqrt{x_1} + \sqrt{x_2}\geq 1$. Let $b=\lfloor {x_2}^{-0.5} \rfloor$.
We may assume that $x_2\leq 1/4$ (otherwise $\sqrt{x_1}+\sqrt{x_2}\geq 1$),
and therefore $b\geq 2$.
Hence
 \begin{multline*}
  \sum_{i=b+1}^\infty x_i \leq \sum_{i=b+1}^\infty
 i^{-2} \leq x_2(x_2^{-0.5}-b) + \int_{x_2^{-0.5}}^\infty
 z^{-2} dz= \\ x_2(x_2^{-0.5}-b) +  \sqrt{x_2}=
 2\sqrt{x_2} - bx_2 .
 \end{multline*}
So,
 \begin{equation*}
x_1=1 - \sum_{i=2}^\infty x_i \geq 1 - (b-1)x_2 - (
2\sqrt{x_2} - bx_2 )= 1- 2\sqrt{x_2} +x_2 = (1-\sqrt{x_2})^2.
 \end{equation*}
That is
$\sqrt{x_1} + \sqrt{x_2}\geq 1$, as needed.
\end{proof}

\begin{proof}[Proof of Lemma~\ref{lem:adv-all-cr}.]
We will use in the proof some elementary probabilistic arguments. For the sake
of completeness, we include their proofs in the appendix.
We derive an upper bound on Formula~\eqref{eq:phase-cost} as follows. Fix
$\delta_1 \in [0,1]$, and denote by $Y$ the event ``$\exists i\geq 2,\
X_i/r_i \leq (1-\delta_1)\mu / r_1$", {\ie}, one of the points in
$\{v_2,\ldots, v_n\}$ has a local cost of at most $(1-\delta_1)\mu / r_1$.
Let $\hat{p}=\Pr[Y]$. We can bound the cost for {\phases} as follows: If $Y$
does not happen, {\phases} can stay in $v_1$, otherwise it moves to $v_i$
with a local cost at most $(1-\delta_1) \mu/ r_1$. Hence its cost is at most
$(1-\hat{p}) E[X_1 | \neg Y] /r_1 + \hat{p} (2+\frac{(1-\delta_1)\mu}{
r_1})$. By Proposition~\ref{prop:depen2} in the appendix,
$E[X_1| \neg Y]\leq E[X_1] =\mu$,
and so we derive the following bound.
\begin{equation} \label{eq:adv-cost-bound}
\E[\cost_{\sopt}(\sigma,v_1)] \leq (1-\hat{p}) \frac{\mu}{r_1} +
\hat{p}(\frac{(1-\delta_1)\mu}{r_1}+2)= \frac{\mu}{r_1} \bigl (1 -
\hat{p} \delta_1) +2 \hat{p}  .
\end{equation}
Assuming  $1\geq \delta_1 \geq 4r_1/\mu$, we have the following bound
\begin{equation*}
\E[\cost_{\sopt}(\sigma,v_1)] \leq  \frac{\mu}{r_1} \bigl (1 -
\hat{p} \tfrac{\delta_1}{2}) =  \beta  .
\end{equation*}

The lower bound on the competitive ratio we achieve is
\begin{equation*}
  r   \geq
 \frac{\mu}{\frac{\mu}{r_1}( 1- \hat{p}  \frac{\delta_1}{2})}
 \geq r_1 (1+ \hat{p}  \frac{\delta_1}{2}).
\end{equation*}

Clearly, we need a lower bound on $\hat{p}$. We define  $p_i= \Pr[X_i \leq
(1-\delta_1) \mu /r_1]$, and analyze the lower bound in two special cases.

\begin{itemize}
\item
In case $r_1=\cdots= r_b$, so $p_1=p_2=\cdots=p_b$. We bound $\hat{p}$ in
terms of $p_1$.
\[ 1-\hat{p}\leq (1-p_1)^{b-1} \leq \exp( - (b-1) p_1)
 \leq   1 - \min \{ \tfrac{1}{4}, \tfrac{1}{2} (b-1) p_1\} .\]
The {first inequality follows from Proposition~\ref{prop:depen} in the appendix,
and
the} last inequality follows since $e^{-\tau} \leq \max \{0.75, 1 - 0.5 \tau
\}$ for $\tau \geq 0$. Thus $\hat{p} \geq \min \{0.25, 0.5 (b-1) p_1 \}$.

To bound $p_1$, we use a lower bound estimate on the tail probability of a
binomial variable. Lemma~\ref{lem:tail-lb} in the appendix states that there
exist constants $\lambda_2\geq 1\geq \lambda_1>0$ such that, $p_1 \geq
\lambda_1 e^{- \lambda_2 \delta_1^2 \mu }$, provided that $\mu \geq 4$. Thus
$\hat{p} \geq \min \{0.25, 0.5 (b-1) \lambda_1 e^{- \lambda_2 \delta_1^2 \mu
}\}$.

Fix $\tilde{\mu}= \frac{16 r_1^2 \lambda_2}{\ln b}$. Note that
$\tilde{\mu}\geq 4$, since $r_1\geq (\ln b)/4$. We want to set $\mu \approx
\tilde{\mu}$, however, we need to maintain $m=n \mu\in \mathbb{N}$, so we
choose $\mu=\lceil \tilde{\mu} \rceil \leq \frac{5}{4}\tilde{\mu}$. In order
to satisfy the constraint on $\delta_1$, we choose
$\delta_1=\frac{4r_1}{\tilde{\mu}}=\frac{\ln b}{4 \lambda_2 r_1}$, so $1\geq
\delta \geq \tfrac{4r_1}{\mu}$.

Since $\delta_1= \sqrt{\frac{\ln b}{\tilde{\mu} \lambda_2}}$, we have
$\hat{p}\geq 0.5\lambda_1 e^{-\frac{5}{4}}\geq \lambda_1/8$. Thus, the lower
bound we show is
\[ r_1+ \hat{p} \tfrac{\delta_1}{2} r_1\geq
 r_1+ \tfrac{\lambda_1}{8} \tfrac{\ln b}{8 \lambda_2 r_1} r_1= r_1+
 \tfrac{\lambda_1}{64{\lambda_2}} \ln b \geq \lambda\ln ( b e^{\lambda^{-1}
 r_1}),
 \]
for $\lambda \leq \tfrac{\lambda_1}{64{\lambda_2}}$. Note that $\beta\leq
\frac{\mu}{r_1} \leq 20\lambda_2 r_1$.

\item
In case $b=2$, let $\delta_2 \in [0,1]$ such that
$(1-\delta_1)\frac{\mu}{r_1}= (1-\delta_2)\frac{\mu}{r_2}$. We fix
$\delta_1=\frac{r_1-r_2+ (20\lambda_2)^{-1}}{r_1}$, so
 \[ \delta_2= \tfrac{r_1-r_2}{r_1}+ \delta_1 \tfrac{r_2}{r_1}
    = \tfrac{r_1-r_2}{r_1}+ \tfrac{r_1-r_2+(20\lambda_2)^{-1}}{r_1} \tfrac{r_2}{r_1}
    \leq 2 \tfrac{r_1-r_2+(20\lambda_2)^{-1}}{r_1}.
 \]
In order to satisfy the constraint on $\delta_1$, we choose
$\mu=\lceil\tilde{\mu}\rceil$, where $\tilde{\mu}=
\frac{4r_1}{\delta_1}=\frac{4r_1^2}{r_1-r_2+(20\lambda_2)^{-1}}$, so $\mu\leq
\frac{5}{4} \tilde{\mu}$. In this case, by applying Lemma~\ref{lem:tail-lb},
 \[ \hat{p}=p_2\geq \lambda_1 e^{- \lambda_2 \delta_2^2 \mu}
    \geq \lambda_1 e^{-16 \frac{5}{4}\lambda_2 (r_1-r_2+(20\lambda_2)^{-1})}=
    \tfrac{\lambda_1}{e} e^{-20 \lambda_2 (r_1-r_2)}.
 \]
Assuming $\lambda\leq \tfrac{\lambda_1}{40 e\lambda_2}$, the lower bound we
show is
 \begin{multline*}
 r_1+ \hat{p} \tfrac{\delta_1}{2} r_1\geq
   r_1+ \tfrac{\lambda_1}{e} e^{-20 \lambda_2 (r_1-r_2)}
   \tfrac{r_1-r_2+(20\lambda_2)^{-1}}{2r_1}r_1
   \geq r_1 + \tfrac{\lambda_1}{40 e \lambda_2} e^{-20 \lambda_2 (r_1-r_2)} \geq \\
   r_1+ \lambda e^{\lambda^{-1} (r_2-r_1)} \geq
   r_1+ \lambda \ln \bigl( 1+ e^{\lambda^{-1} (r_2-r_1)} \bigr ) =
    \lambda \ln \bigl(e^{\lambda^{-1} r_1} + e^{\lambda^{-1} r_2} \bigr ).
 \end{multline*}
 Note that $\beta \leq \frac{\mu}{r_1}\leq 100 \lambda_2 r_1$.
\end{itemize}
In the general case, let $\rho=\lambda /2$, $n_i=e^{\rho^{-1} r_i}$, and
$n=\sum_{i=1}^b n_i$. Applying Proposition~\ref{prop:binary/balanced}, we
get one of the following two possible cases:
\begin{itemize}
\item $\exists \ell$ such that $\ell \sqrt{n_\ell} \geq \sqrt{n}$.
Note that $\min\{\frac{X_1}{r_\ell},2+\min_{i:\ell\geq i\geq 2}
\frac{X_i}{r_\ell}\}$ is an upper bound on Formula~\eqref{eq:phase-cost}.
Thus, our lower bound for $\ell$ equal cost ratios ($=r_\ell$) applies here,
and we get a lower bound of
  \[ \lambda \ln (\ell e^{\lambda^{-1} r_\ell})= \lambda \ln(\ell
  \sqrt{n_\ell})\geq \lambda \ln \sqrt{n}= \rho \ln n.
  \]
\item $\sqrt{n_1}+ \sqrt{n_2} \geq \sqrt{n}$. Again,
$\min\{\frac{X_1}{r_1},2+ \frac{X_2}{r_2}\}$  is an upper bound on
Formula~\eqref{eq:phase-cost}. Thus, our lower bound for $b=2$ applies here,
and we get a lower bound
\[  \lambda \ln \Bigl(e^{\lambda^{-1} r_1} + e^{\lambda^{-1} r_2} \Bigr )=
    \lambda \ln (\sqrt{n_1} +\sqrt{n_2} )\geq
     \lambda \ln \sqrt{n}= \rho \ln n.
     \]
\end{itemize}
We conclude that the claim is proved with the constants
$\lambda_3=100\lambda_2$, and $\rho= \frac{\lambda_1}{2\cdot
64e\lambda_2}$.
\end{proof}

Still, Lemma~\ref{lem:adv-all-cr} requires $r_1\geq \tfrac{1}{4} \ln b$. For
a small $r_1$ we use the standard (fair) MTS lower bound.{\footnote{The
adversary of Lemma~\ref{lem:adv-all-cr} actually works for $r_1\leq \tfrac{
\ln b}{4}$ as well, by choosing $\mu \approx \ln b$, $\delta=1$, and a
simple bound of $\Pr[X=0]\geq 4^{-\mu}$. We choose to present this lower
bound using a different adversary, since the analysis is simpler, and the
bound is better.}
\begin{lemma} \label{lem:fair-mts}
For a UMTS $U=(\mathcal{U}^{\Delta}_b;1,\ldots,1)$  there exists an
$(\frac{H_b}{2} ,2; 1,1,\ldots,1)$-discrete adversary, where $H_n=
\sum_{i=1}^b i^{-1}$.
\end{lemma}
\begin{proof}
Without loss of generality, assume $\Delta=1$. The sequence is determined by
a random permutation $\pi$ of the points in the space. Then,
$\sigma=\tau_1\, \tau_2\, \cdots \tau_b$, where $\tau_i= (v_{\pi(1)},1 )
(v_{\pi(2)},1 ) \cdots (v_{\pi(i)},1 )$.

Obviously, {\sopt}'s cost is at most $2$, because it can move at the
beginning of $\sigma$ to $v_{\pi(b)}$, and return at the end of $\sigma$. The
expected cost of the online on the other hand is at least $\frac{1}{b-i+1}$
in $\tau_i$, and thus at least $\sum_{i=1}^b i^{-1}$ in $\sigma$.
\end{proof}

\begin{lemma} \label{lem:adv-composed}
There exist constants $\lambda_3, \rho >0$ satisfying the following. Given a
UMTS $U=(\U^\Delta_b;r_1,\ldots,r_b)$, with  $r_1\geq r_2 \geq \cdots r_b
\geq 1$, and $(n_i)_i$ satisfying $r_i = \rho (1+ \ln n_i)$, there exists an
$(r,\beta ;\alpha_1,\ldots,\alpha_b)$-discrete adversary such that $r \geq
\rho (1+\ln (\sum_i n_i))$, $\beta \leq \lambda_3 r_1$, and $\min_i \alpha_i
\geq r_b^{-1}$.
\end{lemma}
\begin{proof}
Let $\rho \leq 1/4$ and $\lambda_3 \geq 2$ be the constants from
Lemma~\ref{lem:adv-all-cr}. If $r_1\geq \tfrac{\ln b}{4}$ then the claim
follows from Lemma~\ref{lem:adv-all-cr}.

For $r_1 \leq \frac{\ln b}{4}$, we use the adversary from
Lemma~\ref{lem:fair-mts}. Thus,
\[
r  \geq  0.5 H_b \geq r_1 + \tfrac{\ln b}{4}
 \cfullv{ \geq  \rho(1+\ln n_1) + \rho \ln b}
 \geq \rho (1+ \ln (n_1 b)) .
\]
Also, $\beta=2 \leq \lambda_3 r_1$ and $\min_i \alpha_i=1 \geq r_b^{-1}$.
\end{proof}

\Section{Combining Adversaries on HSTs} \label{sec:hst-lb}

In this section we prove a lemma for combining adversaries for subspaces
using the discrete adversary of Lemma~\ref{lem:adv-composed} as the
combining adversary. We then construct adversaries for HSTs by inductively
combining adversaries for subtrees. When attempting to combine $(r,\beta)$
adversaries, we still have the following problem. The adversary of
Lemma~\ref{lem:adv-composed} can only use specific task sizes, but the tasks
we have from subtrees' adversaries are not necessarily of these sizes. Our
solution is to inductively maintain ``flexible" adversaries that can
generate lower bound sequences with associated optimal cost of value that
may vary arbitrarily in a specified range.

\begin{definition}
Given a UMTS $U=(M;r_1,\ldots,r_b)$ an $(r,\beta ;\eta)$-\emph{flexible}
adversary for $\eta \in [0,1]$ is defined as a collection $\A$ of
$(r,{\beta'})$-adversaries, for all $\beta' \in [{\eta}{\beta}, \beta]$.
\end{definition}

\begin{definition}
Given a UMTS $U$, an
$(r,\beta;\eta;\alpha_1,\ldots,\alpha_b)$-\emph{flexible discrete adversary}
is a collection $\A$ of discrete adversaries for $U$ such that $\forall
\beta'\in [\eta \beta, \beta], \exists (\alpha'_i )_i$ such that
$\alpha'_i\geq \alpha_i$ and $\A$ includes an
$(r,\beta';\alpha'_1,\ldots,\alpha'_b)$-discrete adversary. Obviously, $\A$
is an $(r,\beta ;\eta)$-flexible adversary.
\end{definition}

We start by showing how to transform a discrete adversary into a
flexible discrete adversary with only a small loss in the lower
bound obtained.

\begin{lemma} \label{lem:discrete-flexible}
Denote the UMTSs $U_s=(M;r_1,\ldots,r_b;s)$ with
$\Delta(M)=\Delta$, and assume there exists an $(r, \eta \beta;
\alpha_1,\ldots,\alpha_b)$ discrete adversary $\D_\eta$ for
$U_\eta$, then there exists $(r,\beta; \eta ;\alpha_1,\ldots,
\alpha_b)$ flexible discrete adversary  $\A$ for $U_1$.
\end{lemma}
\begin{proof}
Denote by $U_{\eta,\alpha}=(\alpha M;r_1,\ldots,r_b; \eta)$, (so
$U_\eta=U_{\eta,1}$). $\D_\eta$ is $(r,\eta \beta;
\alpha_1,\ldots,\alpha_b)$ discrete adversary for $U_\eta$.
Observation~\ref{obs:scaling} implies the existence of $(r, \eta
\beta; \alpha_1,\ldots,\alpha_b)$ discrete adversary
$\D_{\eta,{\alpha}^{-1}}$ for $U_{\eta,{\alpha}^{-1}}$ that
replaces each task $(v_i,\alpha_i\Delta)$ of $\D_\eta$ with
$(v_i,\alpha_i \alpha^{-1} \Delta)$.

Consider the adversary $\D_{\eta,\alpha^{-1}}$, for $\alpha\in
[\eta, 1]$ when applied to $U_1$,
\begin{multline*}
\min _{u_0\in M} \E_ {\sigma \in \D_{\eta,\alpha^{-1}}} [
\cost_{\sopt_{U_1}}(\sigma,u_0) ]
 \cconfv{ \\ }
\leq \min _{u_0\in M} \E_ {\sigma \in \D_{\eta,\alpha^{-1}}} [
\cost_{\sopt_{U_{\eta,\alpha^{-1}}}}(\sigma,u_0) ] \leq \eta \beta
\alpha^{-1} \Delta .
\end{multline*}
The first inequality follows since the distances in
$U_{\eta,\alpha^{-1}}$ are larger than in $U_1$.

On the other hand, for any online algorithm $A_{U_1}$ for $U_1$,
consider $A_{U_{\eta,\alpha^{-1}}}$ the simulation of $A_{U_1}$ on
${U_{\eta,\alpha^{-1}}}$. The moving costs for online algorithms
in ${U_{\eta,\alpha^{-1}}}$ are smaller than in $U_1$, since
$\eta\alpha^{-1}\leq 1$. So we have,
 \begin{multline*}
\min _{u_0\in M} \E_ {\sigma \in \D_{\eta,\alpha^{-1}}} [
  \cost_{A_{U_1}}(\sigma,u_0) ]
 \cconfv{ \\ }
\geq \min _{u_0\in M} \E_ {\sigma \in \D_{\eta,\alpha^{-1}}} [
\cost_{A_{U_{\eta,\alpha^{-1}}}}(\sigma,u_0) ] \geq r \eta \beta
\alpha^{-1} \Delta .
  \end{multline*}

Hence, $\D_{\eta,\alpha^{-1}}$ is $(r,\eta \alpha^{-1} \beta )$
adversary for $U_1$. Note that for $v_i$, $\D_{\eta,\alpha^{-1}}$
uses the tasks $(v_i,\alpha_i \alpha^{-1} \Delta)$, and thus it is
actually $(r,\eta\alpha^{-1} \beta ;
\alpha^{-1}\alpha_1,\alpha^{-1}\alpha_2,\ldots,\alpha^{-1}\alpha_b)$-discrete
adversary for $U_1$. Thus $\A=\{\D_{\eta,\alpha^{-1}} \,|\,
\alpha\in [\eta,1] \}$ is an $(r,\beta; \eta ;\alpha_1,\ldots,
\alpha_b)$ flexible discrete adversary for $U_1$.
\end{proof}

\begin{lemma} \label{lem:discrete-to-flexible}
\begin{sloppypar}
The existence of an $(r \eta^{-1}, \eta
\beta;\alpha_1,\ldots,\alpha_b)$-discrete adversary for
$U'=(M; r_1
\eta^{-1},\ldots, r_b \eta^{-1})$ implies the existence of an
$(r,\beta;\eta;\alpha_1,\ldots,\alpha_b)$-flexible discrete adversary for
$U=(M; r_1 ,\ldots, r_b )$.
\end{sloppypar}
\end{lemma}
\begin{proof}
Apply Observation~\ref{obs:w/o-dr} to deduce that the same
adversary is an  $(r, \eta \beta;\alpha_1,\ldots,\alpha_b)$
discrete adversary for $U''=(M;r_1,\ldots,r_b;\eta)$ and then
apply Lemma~\ref{lem:discrete-flexible} on $U''$ to get an
$(r,\beta;\eta;\alpha_1,\ldots,\alpha_b)$ flexible discrete
adversary for $U=(M; r_1 ,\ldots r_b;1 )$.
\end{proof}

\begin{corollary} \label{coro:con-flex}
Given a UMTS $U=(\U^\Delta_b;r_1,\ldots,r_b)$, $r_1\geq \cdots \geq r_b \geq
1$, and $(n_i)_i$ satisfying $r_i = 0.5 \rho (1+\ln n_i)$, there exists an
$(r,\beta ;0.5 ; \alpha_1,\ldots,\alpha_b)$-flexible discrete adversary such
that $r \geq 0.5 \rho (1+ \ln \sum_i n_i)$, $\beta \leq 4 \lambda_3 r_1$,
and $\min_i \alpha_i \geq 0.5 r_1^{-1}$.
\end{corollary}
\begin{proof}
Fix $\eta=0.5$.  Let $\bar{r}_i= r_i \eta^{-1}$. By
Lemma~\ref{lem:adv-composed} we have an
$(\bar{r},\bar{\beta};{\alpha}_1,\ldots,{\alpha}_b)$ discrete adversary for
$(\U^\Delta_b;\bar{r}_1,\ldots,\bar{r}_b)$, satisfying $\min_i \alpha_i \geq
\bar{r}_1^{-1}$, and $\bar{\beta}\leq \lambda_3  \bar{r}_1$. By
Lemma~\ref{lem:discrete-to-flexible}, there exists
$(r,\beta;\eta;{\alpha}_1,\ldots,{\alpha}_b)$ flexible discrete adversary for
$U$, where $r= \eta \bar{r}$ and $\beta=\bar{\beta}\eta^{-1}$. Since
$\bar{r}_i= \eta^{-1} r_i = \eta^{-1} \eta \rho (1+\ln n_i)$, by
Lemma~\ref{lem:adv-composed},
 \( r= \eta \bar{r} \geq \eta \rho \ln (1+\sum_i n_i) \).
\end{proof}

Next we show how to combine flexible adversaries.

\begin{lemma}[Combining Lemma] \label{lem:combine}
Let $U=(M;\bar{r}_1,\ldots, \bar{r}_n)$ be an UMTS, where $M$ is a
$k$-HST metric space of diameter $\delta$ on $n$ points, and
denote the root vertex of the HST by $u$. Let $(M_1, M_2, \ldots,
M_b)$ be the partition of $M$ to subspaces corresponding to the
children $u$.

Let $U_j$ be the UMTS induced by $U$ on $M_j$. Assume that for each $j\in
\{1,\ldots,b\}$ there exists $({r}_j,{\beta}_j;\eta)$-flexible adversary
$\A_j$ for $U_j$. Let $\hat{U}=(\U^{\Delta}_b;{r}_1,\ldots,{r}_b)$ be the
``combining UMTS". Assume there exists an
$(r,\beta;\eta;\alpha_1,\ldots,\alpha_b)$-flexible discrete adversary
$\hat{\A}$ for $\hat{U}$.

If  $k \ge  \frac{\eta}{1-\eta} \max_j \frac{{\beta}_j}
{\alpha_j}$, then there exists a $(r,\beta;\eta)$-flexible adversary $\A$
for $U$.
\end{lemma}
\begin{proof}
We fix $\beta'\in [\eta \beta , \beta]$, and the goal is to construct
$(r,\beta')$ adversary $\D$ for $U$.

We start by choosing a $(r,\beta';\alpha'_1,\ldots,\alpha'_b)$-discrete
adversary $\hat{D}$ from $\hat{\A}$. Denote $\Delta=\Delta(M)$, and
$\Delta_j=\Delta(M_j)$ the diameters of $M$ and $M_j$ respectively. Then,
for each $j$, we choose ${\beta}'_j \in (\eta{\beta}_j,{\beta}_j]$ such that
$t_j=\frac{\alpha'_j \Delta}{{\beta}'_j \Delta_j}$ is a natural number. This
is possible since
 \[
\frac{\alpha'_j \Delta}{\eta {\beta}_j \Delta_j} - \frac{\alpha'_j
\Delta}{{\beta}_j \Delta_j}  \geq \frac{\alpha_j \Delta}{\eta {\beta}_j
\Delta_j} - \frac{\alpha_j \Delta}{{\beta}_j \Delta_j} \geq
\frac{\alpha_j}{\beta_j} \frac{1-\eta}{\eta} k\geq 1.
 \]
Let $\D'_j$ be an $(r_j,\beta'_j)$-adversary from $\A_j$.

We construct a distribution $\D$ on elementary task sequences for $U$ as
follows: first we sample $\hat{\sigma}\in \hat{\D}$; then we replace each
task $(z_j,\alpha'_j \Delta)$ in $\hat{\sigma}$ with $\sigma_j =
\sigma^{(1)}_j\sigma^{(2)}_j\cdots \sigma^{(t_j)}_j$ where each
$\sigma^{(i)}_j$ is independently sampled from $\D'_j$.

Next, we bound ${\sopt}_{U}$. Let $z_{q}$ be the point in $\hat{U}$ that
minimizes $\E_{\hat{\sigma} \in \hat{D}} [\cost_{\sopt_{\hat{U}}}
(\hat{\sigma}, z_q)]$. Let $v_0\in M_{q}$ the point that minimizes
$\E_{\sigma \in ({\D}'_q)} [\cost_{\sopt_{U_q}} ({\sigma}, v_0)]$. Consider
the following offline strategy $B$ for serving $\sigma\in \D$: the algorithm
starts and finishes at $v_0$. $B$ maintains the invariant that if
$\sopt_{\hat{U}}$ is at a point $z_i$ then $B$ is at some point in $M_i$.
Consider some task $(z_j,\alpha'_j \Delta)$ in $\hat{\sigma}$. It is replaced
by sequence $\sigma_j$ as described above. If $\sopt_{\hat{U}}$ moves to a
point different from $z_j$ it incurs a cost of $\Delta$. In this case $B$
moves out of $M_j$ ahead of the task sequence $\sigma_j$ incurring a cost of
$\Delta$ as well. If $\sopt_{\hat{U}}$ is not at $z_j$ then its cost for the
task is 0 and the cost for $B$ on $\sigma_j$ is also 0. Otherwise,
$\sopt_{\hat{U}}$ incurs a cost of $\alpha'_j\Delta$ for the task. In this
case $B$ uses $\sopt_{U_j}$ to serve $\sigma_j$ in $M_j$. The expected cost
of ${\sopt}_{U_j}$ for each subsequence $\sigma^{(i)}_j$ of $\sigma_j$ is at
most ${\beta}'_j \Delta_j$, and therefore the cost of $B$ equals
\[
\min_{u_0 \in M_j} \E_{\sigma_j \in ({\D}'_j)^{t_j}}
[\cost_{\sopt_{U_j}}(\sigma_j,u_0)] \le t_j{\beta}'_j \Delta_j =
\alpha'_j\Delta .
\]
It follows that the expected cost of $B$ for serving $\sigma_j$ is bounded
from above by the cost of $\sopt_{\hat{U}}$ on the task $(z_j,\alpha'_j
\Delta)$. Hence
\[
\cfullv{ \min_{u_0 \in M} \E_{\sigma\in \D}[\cost_{\sopt_U}(\sigma,u_0)]
\leq}  \E_{\sigma\in \D}[\cost_{B}(\sigma,v_0)] 
 \leq \E_{\hat{\sigma}\in
\hat{\D}}[\cost_{{\sopt}_{\hat{U}}}(\hat{\sigma},z_{q})] \leq \beta' \Delta.
\]

It is left to show  a lower bound on online algorithms for $U$. Let $A$ be
an online algorithm for $U$. We can naturally define an online algorithm
$\hat{A}$ for $\hat{U}$ as follows. Consider a distribution on sequences
$\sigma \in \D$ generated as described above from a sequence $\hat{\sigma}
\in \hat{\D}$. Consider a task $(z_j,\alpha'_j\Delta)$ in $\hat{\sigma}$ and
let $\sigma_j$ be the corresponding sequence generated above for $M_j$.
Whenever $A$ moves between subspaces into a point in subspace $M_i$ during
the service of $\sigma_j$, if $i\neq j$ then $\hat{A}$ makes a move to the
corresponding point $z_i$ before serving the task. $\hat{A}$ serves the task
in the last such $z_i$ and finally moves to the point corresponding to the
subspaces in which $A$ ends the service of $\sigma_j$. Obviously, the moving
cost of $\hat{A}$ is bounded from above by the cost $A$ incurs on moves
between subspaces. If $A$ does move between subspaces during the service of
$\sigma_j$ then $\hat{A}$ incurs zero local cost for the task and therefore
its cost for the task is at most that of $A$ on $\sigma_j$. Otherwise, if
$A$ is in a subspace $M_i$ $i\neq j$ during the entire sequence $\sigma_j$
then the cost of $\hat{A}$ for the task is 0. If $A$ is in $M_j$ during
$\sigma_j$ then we have
 \[
\min_{u_0 \in M_j} \E_{\sigma_j \in ({\D}'_j)^{t_j}}
 [\cost_{A}(\sigma_j,u_0)] \ge t_j \cdot r_j {\beta}'_j \Delta_j =
r_j\alpha'_j\Delta ,
 \]
which is the cost for $\hat{A}$. It follows that in all cases the expected
cost of $A$ on $\sigma_j$ is at least the cost of $\hat{A}$ on the task.
Thus, we get that for any online algorithm $A$,

 \begin{equation*}
\min_{u_0 \in M} \E_{\sigma\in \mathcal{D}}[\cost_{A}(\sigma,u_0)] \geq
\cconfv{ \\ }
 \min_{z_0 \in \hat{M}} \E_{\hat{\sigma}\in \hat{D}}[
\cost_{\hat{A}}(\hat{\sigma},z_0)] \geq r\beta'\Delta . \qquad \qed
 \end{equation*}
\renewcommand{\qed}{}
\end{proof}

\begin{proof}[Proof of Theorem~\ref{thm:hst-lb}.]
Fix constants $ c_2=0.5 \rho $, $c_3=4 \lambda_3$, and $c_1=2  c_3$.
Consider an arbitrary $(c_1 (1+\ln N)^2)$-HST on $N$ points. We construct by
induction on the height of a subtree $T_u$ rooted at $u$, a $(r _u, \beta
_u;0.5)$-flexible adversary for a subtree with $n_u$ leaves, such that $r_u
\geq \max\{1,c_2 (1+\ln n_u)\} $, and $\beta_u\leq c_3 (1+\ln n_u)$.

The base case are trees on of height $1$ for which we can apply the
adversary of Lemma~\ref{lem:fair-mts} with $r_i=1$.

For height larger than one, assume an internal vertex $u$ of the HST has $n$
points in its subspace, and $b$ children. Inductively, assume that each
$T_i$, a tree rooted at the children of $u$, has $(r_i,\beta_i;0.5)$ flexible
adversary, such that $\beta_i \leq c_3 (1+\ln n_i)$ and $r_i\geq \max\{1, c_2
(1+\ln n_i)\}$. Note that $(r,\beta; \eta)$ flexible adversary implies
$(r',\beta;\eta)$ flexible adversary for $r' \leq r$, and therefore we may
assume that $r_i= \max\{1, c_2(1+ \ln n_i)\}$.

We use the flexible discrete adversary from
Corollary~\ref{coro:con-flex} as the combining adversary in
Lemma~\ref{lem:combine}. Here $\beta_i / \alpha_i \leq 2 r_i \beta
_i \leq 2 \max\{c_2(1+\ln n_u),1\} c_3 (1+\ln n_u)\leq c_1 (1+\ln
N)^2$, and thus we get an $(r,\beta; 0.5)$ flexible adversary for
$T_u$ with $r \geq c_2 (1+\ln n_u)$, and $\beta \leq 4 \lambda_3
\max_i r_i \leq c_3 (1+\ln n_u)$.
\end{proof}

\begin{corollary} \label{cor:mesh-lb}
The randomized competitive ratio of the MTS problem in $n$-point
$\ell$-dimensional mesh is $\Omega  (\frac{\log n}{ \log \log n} )$.
\end{corollary}
\begin{proof}
Combining Lemma~\ref{lem:mesh} with Lemma~\ref{lemma:hst-subtree-approx},
using $k=\Theta(\log^2 n)$, we deduce that the mesh contains a subspace of
size $n^{\Omega(\frac{1}{\log \log n})}$ that $O(1)$ approximates a
$\Omega(\log^2n)$-HST. Next we apply the lower bound of
Theorem~\ref{thm:hst-lb} on that HST.
\end{proof}

\Section{Lower bounds for $K$-server} \label{sec:servers}

Theorem~\ref{thm:main} also implies a lower bound for the $K$-server problem.
This follows from the following general reduction from the MTS problem on an
$n$ point metric space to the $(n-1)$-server problem on the same metric
space.

 \newcommand{\lcost}{{\text{lcost}}}
 \newcommand{\mcost}{{\text{mcost}}}

\begin{lemma} \label{lem:MTS-servers}
An\/ $r$-competitive randomized algorithm for the $(n-1)$-servers problem
on an
$n$-point metric space against oblivious adversaries implies a $(2r+1)$
upper bound on the randomized competitive ratio for MTS on the same metric
space.
\end{lemma}
\cconfv{ \begin{proof} Omitted. \end{proof}}
\begin{fullv}
\begin{proof}
We will prove the implication only for MTS problems in which the
tasks are elementary. For the purpose of establishing a lower
bound for the $K$-server this is enough since our lower bound for
MTS uses only elementary tasks. However, there is also a general
reduction \cite{BBBT97} from an upper bound for any tasks to an
upper bound for elementary tasks.

Given a metric space $M$ on $n$ points with metric $d$ and diameter $\Delta$,
denote by $S$ the $(n-1)$-servers problem on $M$ and by $T$ the MTS problem
on $M$. For a request sequence $\sigma$ in $S$, and point $i\in M$, we denote
by $w^S_{\sigma}(i)$ the optimal offline cost for servicing $\sigma$ and end
without a server in $i$. Similarly for task sequence $\tau$ in $T$ we denote
by $w^T_\tau(i)$ the optimal cost for servicing $\tau$ and ending in state
$i$ (these are called \emph{work functions}). Note that for any
$\tau,\sigma,i,j$,  $w^S_\sigma(i) - w^S_\sigma(j) \leq d(i,j)$ and
$w^T_\tau(i) - w^T_\tau(j) \leq d(i,j)$.

Given a randomized algorithm $A_S$ for $S$, we construct an
algorithm $A_T$ for $T$. $A_T$ transforms a task sequence $\tau$
into a sequence $\sigma$ for $S$ as follows. Assume the sequence
is $\tau'=\tau e$ where $\tau$ has been already transformed into
$\sigma$. $A_T$ transforms the elementary task $e=(i,\delta_i)$
using the following rule. If
\begin{equation} \label{eq:trans-eq}
w^T_\tau(i)+\delta_i \geq \min_{j:j\neq i} w^T_\tau(j)+d(i,j),
\end{equation}
it gives a request for $i$ in $S$, otherwise no request is given.
$A_T$ simulates $A_S$ and maintains its state in the point where
$A_S$ does not have a server.\footnote{Without loss of generality,
we may assume that no two servers of $A_S$ are at the same point.}
Note that the request sequence $\sigma$ was constructed oblivious
to the random bits of $A_S$, and thus $\E [\cost_{A_S}(\sigma)]
\leq r \cost_{\opt_S}(\sigma) + C$.

Next, we prove by induction on the sequence that for any $\tau$
and any $i$, $w_\sigma^S(i) \leq w_\tau^T(i)$.  For $\tau=\e$, it
is obvious that for all $i$, $w^S_\e(i)=w^T_\e(i)$. Tasks in $T$
that do not generate tasks in $S$, obviously maintain the
inductive invariant. Otherwise, let $e=(l,\delta_l)$ be a task in
$T$ that generates a request $e'$ in $S$ for $l$.  $w^S$ has the
following update rules. In point $l$,
 \begin{equation*}
w^S_{\sigma e'}(l) = \min_{j:j\neq l} (w^S_{\sigma}(j)+
d(l,j))\leq
 \min_{j:j\neq l}(w^T_{\tau}(j)+ d(l,j)) = w^T_{\tau e}(l) .
 \end{equation*}
 The last equality follows from \eqref{eq:trans-eq}.
For $i\neq l$, $w^S_{\sigma e'}(i)=w^S_\sigma(i)\leq
w^T_\tau(i)\leq w^T_{\tau e}(i)$. Therefore
$\cost_{\opt_S}(\sigma) \leq \cost_{\opt_T}(\tau)$.

Denote by $\lcost_A(\tau)$ and $\mcost_A(\tau)$ the local cost and
the movement cost of algorithm $A$ on sequence $\tau$. Since $A_T$
moves similarly to $A_S$, $\mcost_{A_T}(\tau) =
\mcost_{A_S}(\sigma)$. To bound the local cost of $A_T$, we prove
that
\begin{equation} \label{eq:costA}
\lcost_{A_T}(\tau) \leq \mcost_{A_T}(\tau) + w^T_{ \tau}(i_c),
\end{equation}
where $i_c$ is the current state of $A_T$. Consider a task
$e=(i,\delta_i)$. If $A_T$ was not in a state $i$, no local cost
was generated. If $A_T$ was in a state $i$ and did not move in
response to task $e$, its local cost is $\delta_i$. On the other
hand, for any $j$, $w^T_\tau(i)+\delta_i - w^T_\tau(j) \leq
d(i,j)$, so $w^T_{\tau e}(i)=w^T_{\tau }(i)+\delta_i$, hence
Eq.~\eqref{eq:costA} is maintained. If $A_T$ moves to state $j$
then its local cost is $0$. In this case the right side of
Eq.~\eqref{eq:costA} is changed by
$d(i,j)+w^T_\tau(j)-w^T_\tau(i)\geq 0$, and therefore
Eq.~\eqref{eq:costA} is maintained. We conclude that
$\cost_{A_T}(\tau) \leq 2 \cost_{A_S}(\sigma) +
\cost_{\opt_T}(\tau)+ \Delta$. To summarize
 \begin{multline*}
 \E [\cost_{A_T}(\tau)] \leq  2 \E[\cost_{A_S}(\sigma)] + \cost_{\opt_T}(\tau)+\Delta \\
\leq
 2r\cost_{\opt_S}(\sigma) + \cost_{\opt_T}(\tau) +\Delta+ 2C
 \cconfv{\\ } = (2r +1)\cost_{\opt_T}(\tau) +C' ,
 \end{multline*}
where $C'=2C+\Delta$ is a constant.
\end{proof}

We remark that the technique of Lemma~\ref{lem:MTS-servers} can
also be applied to deterministic algorithms, but not directly to
randomized algorithms in the \emph{adaptive online adversary
model}~\cite{BBKTW94}. In \cite{MMS90}, a different reduction from
the MTS problem to the servers problem is given. Their reduction
applies to all adversary models and is more efficient. However, it
reduces an MTS problem to a servers problem in a \emph{different
metric space}, and therefore inappropriate for our purposes.
\end{fullv}

When applying Lemma~\ref{lem:MTS-servers} on
Theorem~\ref{thm:main} we deduce
following.\footnote{A direct way to argue
Theorem~\ref{thm:servers} without using
Lemma~\ref{lem:MTS-servers} is to observe that the adversary in
the proof of Theorem~\ref{thm:hst-lb} uses tasks that if replaced
with task size infinity will increase {\sopt}'s cost by at most a
factor of two.}

\begin{theorem} \label{thm:servers}
The randomized competitive ratio against oblivious adversaries of
the $K$-server problem on a metric space with more than $K$ points
is\/ $\Omega(\log K/ \log^2\log K)$.
\end{theorem}

Using Corollary~\ref{cor:mesh-lb} we have

\begin{corollary} \label{cor:servers-meshes}
The randomized competitive ratio against oblivious adversaries of
the $K$-server problem on $h$-dimensional mesh with more than $K$ points
is $\Omega(\log K/ \log \log K)$.
\end{corollary}
\begin{proof} [Proof sketch.]
Let $M$ be an $h$-dimensional mesh, $[s]^h$.
We first remove points from $M$ to obtain a maximal sub-mesh, $M'$, of $M$
of size $m \leq K$. It easy to observe that $m \geq \sqrt{K}$.
It follows from Lemma~\ref{lem:MTS-servers} and Corollary~\ref{cor:mesh-lb}
that $M'$ has a lower bound of $\Omega(\log m/\log\log m)$
for $m-1$ servers.
To get a lower bound for $M$ we pick $K-m+1$ points not in $M'$ and
modify the adversary for $M'$ by inserting repeated requests to these points
between its original requests to make sure that $K-m+1$ servers will have
to stay at these points.
\end{proof}

For $n\gg K$, it is possible to get a better lower bound.

\begin{theorem} Fix a constant $\e>0$. Then for any $K$
and any metric space $M$ on $n \geq K^{\log ^\e K}$ points, the $K$-server
problem on $M$ has a lower bound of\/ $\Omega(\log K)$ on the competitive
ratio for randomized online algorithms against oblivious adversaries.
\end{theorem}
\begin{proof} [Proof sketch.]
Assume $K$ is large enough. Let $f=K^{\log ^\e K}$. We take an
arbitrary subspace with $f$ points. Using
Theorem~\ref{thm:gen-khst-subspace} with $\beta=\log^{\e/2} K$,
$\ell=2$, and $k=\Theta(\log ^2 K)$, we find a subspace that
$O(\log_\beta \log f)= O(\e^{-1})$ approximates a $k$-HST and has
$f^{(\beta \lceil \log k\rceil)^{-1}}>K$ points. We further delete
arbitrary points from this sub-space to get exactly $K+1$ points.
From Theorem~\ref{thm:hst-lb} we have a lower bound of
$\Omega(\log K)$ MTS in this space. We conclude the claim by using
Proposition~\ref{prop:metric_approx_lb} and
Lemma~\ref{lem:MTS-servers}.
\end{proof}

\Section{Additional Ramsey-type Theorems} \label{sec:additional Ramsey}

In this section we prove additional Ramsey-type theorems, and
relate our constructions to those of \cite{BFM86,KRR94,BKRS00}.
In a subsequent paper \cite{BLMN-dichotomy} these metric Ramsey problems are
further studied, and tight bounds are given.

\begin{definition}
A vertex $u$ in a rooted tree is called \emph{balanced} if the difference
between the number of leaves of any two subtrees rooted at $u$'s children,
is at most one. The following is a decreasing hierarchy of HST subclasses.

\begin{enumerate}
\item A \emph{``binary/balanced" $k$-HST} is a $k$-HST with the
property that every internal vertex is either balanced or has at
most two children.

\item A \emph{``binary/uniform" $k$-HST} is a $k$-HST with the property that
every internal vertex $u$ either has at most two children or all
its children are leaves.

\item A \emph{``BKRS" $k$-HST}
is a ``binary/uniform" $k$-HST such that an internal
vertex with exactly two children is either balanced or one of the children is
a leaf.

\item A \emph{``BFM" HST}
is a 1-HST whose underlying tree is binary and for each vertex at most one
child is not a leaf.

\item A \emph{``KRR" $k$-HST}, for $k>1$,
is either a uniform space or a $k$-super increasing metric space,
where a $k$-super increasing space is a $k$-HST in which every
internal vertex has at most two children, and at most one of them
is not a leaf.
\end{enumerate}
\end{definition}

Bourgain {\etal}~\cite{BFM86}, Karloff {\etal}~\cite{KRR94} and
Blum {\etal}~\cite{BKRS00} essentially prove
the following Ramsey-type theorems.

\begin{theorem}\label{thm:special-hst}
For any $k\geq 4$ and any metric space $M=(S,d)$ on $n$ points:
\begin{enumerate}
\item (\cite{BFM86}) There exists a subspace $S'\subseteq S$ such that
$|S'|\geq C(\e){\log n}$ and $(S',d)$ is
$(1+\e)$-approximated by a ``BFM" HST.\footnote{
This is statement is only implicit in \cite{BFM86}. They are interested
in embedding a subspace inside $\ell_2$. Embedding in $\ell_2$ is achieved by
observing that a ``BFM" HST is isomorphic to a subset of $\ell_2$.}

\item (\cite{KRR94}) There exists a subspace $S'\subseteq S$ such that $|S'| =
\Omega(\frac{\log n}{\log \log n})$ and $(S',d)$ is
$O(k^2)$-approximated by a ``KRR" $k$-HST.\footnote{Using
Lemma~\ref{lem:1-hst}, it is possible to improve the theorem to
$O(k)$ approximation by a ``KRR" $k$-HST.}
\item (\cite{BKRS00}) There exists a sub-space $S'\subseteq S$ such that $|S'| = 2^{\Omega(
\sqrt{\log_k n} - \log ^2 k)}$, and $(S',d)$ is $4$-approximated by a ``BKRS"
$k$-HST.\footnote{The definition of ``BKRS" HST, the
statement of this claim, and its proof are only implicit in \cite{BKRS00}.
In particular, they only consider the case $k= \log^3 n$.}
\end{enumerate}
\end{theorem}

``Binary/balanced" HSTs are of special interest for us. Our lower
bound on the competitive ratio of HST is actually proved for this
class of spaces, with Proposition~\ref{prop:binary/balanced} as
the key argument for applying it on arbitrary HST (see
Lemma~\ref{lem:adv-all-cr}). Here we explicitly construct
``binary/balanced" HSTs.

\begin{lemma}\label{lem:binary-or-balanced HST}
In any HST on $n$ leaves there exist a subset of the leaves of size
$\sqrt{n}$ on which the induced HST is a ``binary/balanced" HST.
\end{lemma}
\begin{proof}
The lemma is proved inductively by applying
Proposition~\ref{prop:binary/balanced} as the inductive argument.
The only issue here is how to maintain subtrees with the same
number of leaves. This is done using a dynamic programming
approach.

Formally, we prove by induction on $h$ that for any rooted tree $T$ with a
root $r$, of height $h$, and with $n$ leaves, and for any $m\in\{0,1,\ldots,
\lceil \sqrt{n}\rceil\}$, $T$ contains a sub-tree rooted at $r$ on $m$
leaves.

For $h=0$ the claim is trivial. Otherwise, let $T_1,\ldots, T_b$
be the subtrees rooted at the children of $r$. Denote by
$n_i=|T_i|$, so $n=\sum_{i=1}^b n_i$. Assume without loss of
generality that $n_1\geq n_2\geq\cdots \geq n_b>0$. If $b=1$ the
claim follows by the inductive hypothesis on $T_1$. Otherwise, fix
an integer $m$, $\lceil \sqrt{n} \rceil \geq m\geq 0$. By
Proposition~\ref{prop:binary/balanced}, one of the following
holds:
\begin{enumerate}
\item $\sqrt{n}\leq \sqrt{n_1} + \sqrt{n_2}$. In this
case we choose integers $m_1\leq \lceil \sqrt{n_1} \rceil$ and $m_2 \leq
\lceil \sqrt{n_2} \rceil$ such that $m=m_1+m_2$. By the inductive hypothesis
there exist $T'_1$, a "binary/balanced" subtree of $T_1$ with $m_1$ leaves
and $T'_2$ a "binary/balanced" subtree of $T_1$ with $m_2$ leaves. The tree
$T'$ --- rooted at $r$ with the two  subtrees $T'_1$ and $T'_2$ as the
children --- is a ``binary/balanced" tree with $m$ leaves.

\item $\exists \ell \in \{2,\ldots,b\}$ such that $\sqrt{n}
\leq \ell\, \sqrt{n_\ell}$. Let $m'=m/\ell$. Note that $m' \leq
\lceil\sqrt{n_\ell} \rceil$, since
 $m \leq \lceil\sqrt{n} \rceil \leq \ell \lceil\sqrt{n_\ell}
\rceil$. Thus, by the inductive hypothesis, for $i \leq \ell$, it
is possible to extract from $T_i$ trees with $\lfloor m'\rfloor$
and $\lceil m' \rceil$ leaves. By choosing a combination of trees
$T'_i$ of sizes $\lfloor m'\rfloor$ or $\lceil m' \rceil$, it is
possible to get trees $T'_i$ such that
$|T'_i|-|T'_j|\in\{-1,0,1\}$ and $\sum_i |T'_i|=m$. Combining
these subtrees with the root $r$, gives a ``binary/balanced"
subtree $T'$ with $m$ leaves. \qed
\end{enumerate}
\renewcommand{\qed}{}
\end{proof}

Lemma~\ref{lem:binary-or-balanced HST} combined with
Theorem~\ref{thm:gen-khst-subspace}, is the strongest Ramsey-type theorem
presented in this paper.

\begin{theorem} \label{thm:bu-khst-subspace}
For any metric space $M=(V,d)$ on $|V|=n$ points, any $\beta
>1$, any $k>1$, and any $1< \ell \leq k$
there exists a subset $S\subseteq V$, such that $|S|\geq n ^{\frac{1}{2\beta
\lceil \log_\ell k \rceil}}$ and $(S,d)$  $O(\ell \log_\beta \log
n)$-approximates a binary/balanced $k$-HST.
\end{theorem}

\begin{proposition} \label{lem:KRR-HST}
In any $k$-HST on $n$ leaves there exist a subset of the leaves of size
$\Omega(\frac{\log n}{\log \log n})$ on which the induced HST is a ``KRR"
$k$-HST.
\end{proposition}
\begin{proof}
Let $T$ be the given HST and assume it does not have degenerate vertices.
Either $T$ has an internal vertex $u$ with at least $\log n$ children, and
in this case, by taking one descendant leaf from each child of $u$, we get a
uniform space. Otherwise, $T$ must have a vertical path of length at least
$\log_{\log n} n$. Take this path and add for each internal vertex along
the path another child as a leaf. The resulting HST is super-increasing.
\end{proof}

\begin{proposition} \label{prop:BFM-HST}
In any HST on $n$ leaves there exists a subset of the leaves of size
$\Omega({\log n})$ on which the induced metric space  is a ``BFM" HST.
\end{proposition}
\begin{proof}
We first observe that any HST can be transformed into a $1$-HST whose
underlying tree is binary without degenerate vertices. We then take the
longest vertical path $p$ in $S$ --- its length is at least $\log n$ --- and
adjoin for each internal vertex $u$ along $p$, a leaf from the subtree of
the child of $u$ not on $p$.
\end{proof}

\begin{proposition} \label{prop:BKRS}
Given a sequence $n_1\geq n_2\geq \cdots \geq n_b>0$, and
$n=\sum_{i=1}^b n_i$, then $\max\{b, 2 n_2^{\frac{1}{2\sqrt{\log
n}}}, n_1 ^{\frac{1}{2\sqrt{\log n}}} +1 \}\geq
2^{\frac{\sqrt{\log n}}{2}}$.
\end{proposition}
\begin{proof}
Assume that $\max\{b, 2 n_2^{\frac{1}{2\sqrt{\log n}}} \}<
2^{\frac{\sqrt{\log n}}{2}}$. Then
 \[ n_1 \geq n- b n_2 \geq n -  2^{\frac{\sqrt{\log
n}}{2}} \frac{n}{2^{2\sqrt{\log n}}} \geq
 n(1- \frac{1}{2^{\sqrt{\log n}}}). \]
Therefore,
 \[ n_1^{\frac{1}{2\sqrt{\log n}}} \geq n^{\frac{1}{2\sqrt{\log n}}}
  (1- \frac{1}{2^{\sqrt{\log n}}})^{\frac{1}{2\sqrt{\log n}}}\geq
  2^{\frac{\sqrt{\log n}}{2}} (1- \frac{1}{2^{\sqrt{\log n}}}) \geq
  2^{\frac{\sqrt{\log n}}{2}} -1. \quad \qed
\]
\renewcommand{\qed}{}
\end{proof}

\begin{proposition} \label{lem:BKRS-HST}
In any HST on $n$ leaves there exists a subset of the leaves of
size at least $2^{\sqrt{\log n}/2}$  on which the induced HST is a
``BKRS" HST.
\end{proposition}
\begin{proof}
We prove,  by induction on the height of the tree, that for any tree $T$ with
$n$ leaves and for any $m\leq \lceil 2^{\frac{\sqrt{\log n}}{2}} \rceil$,
$T$ contains a ``BKRS" subtree $T'$ with $m$ leaves.

Let $r$ be the root of $T$ and $T_1,\ldots,T_b$ the subtrees rooted at the
children of $r$. Denote $n_i=|T_i|$, and apply Proposition~\ref{prop:BKRS}.

If $b\geq 2^{\frac{\sqrt{\log n}}{2}}$, we construct $T'$ by connecting $r$
to one leaf from each $T_i$, for $1\leq i\leq m $.

If $2 n_2^{\frac{1}{2\sqrt{\log n}}} \geq 2^{\frac{\sqrt{\log n}}{2}}$, then
we construct $T'$ by connecting $r$ to $T'_1$ and $T'_2$, where $T'_1$ is a
subtree of $T_1$ with $\lceil m/2 \rceil$ leaves, and $T'_2$ is a subtree of
$T_2$ with $\lfloor m/2 \rfloor$ leaves.

If $n_1 ^{\frac{1}{2\sqrt{\log n}}} +1 \geq 2^{\frac{\sqrt{\log n}}{2}}$,
then we construct $T'$ by connecting $r$ to $T'_1$ and one leaf from $T_2$,
where $T'_1$ is a subtree of $T_1$ with $m-1$ leaves.
\end{proof}

Proposition~\ref{lem:KRR-HST}, Proposition~\ref{prop:BFM-HST},
 and Proposition~\ref{lem:BKRS-HST}, when
combined with Theorem~\ref{thm:gen-khst-subspace}, give corresponding
Ramsey-type theorems. These results, however, are slightly weaker than
Theorem~\ref{thm:special-hst}, as the approximation factor is $O(\log \log n)$
instead of  a
constant.\footnote{This is when using a constant $\beta$. Alternatively, when
choosing $\beta=\log ^{\e}n$, we get a constant approximation but of slightly
smaller subspaces.} We include them to demonstrate the simplicity of their
proof, when using HST.

We end the section with some impossibility examples. The first one deals with
subspaces of equally spaced points on the line.

\begin{proposition} \label{prop:line-inapprox}
For any $\alpha\geq 1$ there exists $c<1$, such that any subset of
$n$ equally spaced points on the line that is
$\alpha$-approximated by an HST, is of size at most $O(n^c)$.
\end{proposition}
\begin{proof}
Let $M=\{v_1,v_2,\ldots,v_n\}$ be the metric space on $n$ points such that
$d_M(v_i,v_j)=|i-j|$. Let $S\subseteq M$ be a subspace that is $\alpha$
approximated by an HST $T$. We prove by induction on $n$ that $|S|\leq
2(\alpha+1){n}^{c} $, where $c=c(\alpha)<1$ will be chosen later.

Without loss of generality, we may assume that $T$ is a binary
tree without degenerate vertices. Let $n'=\max\{d_M(u,v) | u,v \in
S\}+1 \leq n$. Without loss of generality, assume that
$v_1,v_{n'}\in S$ are the two extreme points in $S$. For $n'\leq
2(\alpha+1)$ the inductive claim is trivially true. Otherwise, let
$u=\lca_T(v_1,v_{n'})$, so $\Delta(u)\geq n'-1$. Denote by $S_1$
and $S_2$ the two subspaces induced by the children of $u$.  Since
for any $v\in S_1$ and $v'\in S_2$, $d_M(v,v')\geq (n'-1)/\alpha$,
we can partition the interval $\{v_1,\ldots, v_{n'}\}$ into
$2\ell+1$ sub-intervals $I_1,\ldots I_{2\ell+1}$, such that for
any $i\in\{1,\ldots, \ell\}$ $|I_{2i}| \geq
\tfrac{n'-1}{\alpha}-1$ and $I_{2i} \cap S= \emptyset$; for $i\geq
0$, $I_{4i+1}\cap S \subseteq S_1$ and $I_{4i+3}\cap S \subseteq
S_2$. Denote by $n_i=|I_{i}|$. Thus $\sum_{i=0}^\ell n_{2i+1} +
\ell (\alpha(n'-1)-1)\leq n'$. The induced HST on $S\cap I_{2i+1}$
$\alpha$-approximates the original distances, and therefore by the
inductive hypothesis $|S\cap I_{2i+1}|\leq 2(\alpha+1) n_{2i+1}^
c$.

Assume $\ell=1$, then $n_1+n_3\leq n'-
(\tfrac{n'-1}{\alpha}-1)\leq n'(1-\tfrac{1}{2\alpha})$, the last
inequality follows since $n'\geq 2(\alpha+1)$. By concavity, the
maximum of $n_1^c +n_3^c$ is reached when
$n_1=n_3\leq(n'(1-\tfrac{1}{2\alpha}))/2$. Thus
\begin{multline*}
|S|\leq 2(\alpha+1)2 \bigl(\frac{n'(1-\tfrac{1}{2\alpha})}{2} \bigr )^c
\leq   2(\alpha+1) 2 \bigl(\frac{(1-\tfrac{1}{2\alpha})}{2}\bigr)^c n'{}^{c}
\leq 2(\alpha+1)n'{}^c.
\end{multline*}
The last inequality follows since it is possible to choose $c<1$ such that $2
\bigl(\frac{(1-\tfrac{1}{2\alpha})}{2}\bigr)^c \leq 1$.

The proof for $\ell>1$ follows by induction on $\ell$. Denote
$J_1=\cup_{i=1}^{2\ell-1}I_i$, $J_2=I_{2\ell}$, and
$J_3=I_{2\ell+1}$. Also denote $N_1=|J_1|$ and $N_3=|J_3|$. By the
inductive hypothesis, $|S\cap J_1|\leq 2(\alpha+1)N_1^c$, and
$|S\cap J_3|\leq 2(\alpha+1) N_3^c$. Applying the argument above,
we conclude that $N_1^c+N_3^c\leq n'{}^c$.
\end{proof}

Next we show examples that prove that Lemma~\ref{lemma:hst-subtree-approx},
Theorem~\ref{thm:special-hst}, Proposition~\ref{lem:KRR-HST},
Proposition~\ref{prop:BFM-HST} and
Proposition~\ref{lem:BKRS-HST} are all essentially tight. Before presenting
the examples we need the following claims.

\begin{proposition} \label{prop:lca-in-HST}
Assume that an HST $T$ is $\ell$-approximated by a $k$-HST $W$, and $\ell<k$.
Then for any four (not necessarily distinct) points $a,b,c,d$ in the space,
 \[ \lca_T(a,b)=\lca_T(c,d) \implies \lca_W(a,b)=\lca_W(c,d). \]
\end{proposition}
\begin{proof}
Assume $\lca_T(a,b)=\lca_T(c,d)$. Denote $u'=\lca_W(a,b)$, and
$v'=\lca_W(c,d)$. Assume for the sake of contradiction that
$u'\neq v'$. First we observe that  $u'$ can not be a proper
ancestor of $v'$, since otherwise $d_W(a,b)>\ell d_W(c,d)$, and
this means that $W$ does not $\ell$ approximates $T$. From the
same reason $v'$ is not a proper ancestor of $u'$. This implies
that $\lca_W(a,c)$ is a proper ancestor of $\lca_W(a,b)$, and so
$d_W(a,c)>\ell d_W(a,b)$, whereas in $T$ it must be that
$\lca_T(a,c)$ is a descendant of $\lca_T(a,b)$, and thus
$d_T(a,c)\leq d_T(a,b)$. Again, this means that $W$ does not
$\ell$-approximate $T$, a contradiction.
\end{proof}

\begin{lemma} \label{lem:iso-HSTs}
Assume that a $k$-HST $T$ is $\ell$-approximated by a $k$-HST $W$. If both
$T$ and $W$ do not have degenerate vertices and $\ell<k$, then the underlying
trees of $T$ and $W$ are isomorphic.
\end{lemma}
\begin{proof}
It is sufficient to show that for any four (not necessarily distinct) points
$a,b,c,d$ in the space, $\lca_T(a,b)=\lca_T(c,d)$ if and only if
$\lca_W(a,b)=\lca_W(c,d)$. This is so since we can define $f:T\rightarrow
W$, by $f(\lca_T(a,b))=\lca_W(a,b)$. It is easy to check that $f$ is well
defined injective and bijective. Also, if $u$ is ancestor of $v$ in $T$,
then  $f(u)$ is an ancestor of $f(v)$ in $W$. To see this, Let $a,b$ two
descendant leaves of $v$ in $T$ such that $\lca_T(a,b)=v$, and let $c$ be a
descendant leaf of $u$ such that $\lca_T(a,c)=\lca_T(b,c)=u$, but then
$\lca_W(a,c)=\lca_W(b,c)$, and this implies that $\lca_W(a,c)$ is an
ancestor of $\lca_W(a,b)$.

In order to prove that $\forall a,b,c,d$, $\lca_T(a,b)=\lca_T(c,d)$ if
and only if $\lca_W(a,b)=\lca_W(c,d)$, we apply
Proposition~\ref{prop:lca-in-HST} in two directions
(and noting that the approximation relation is essentially symmetric,
see the discussion after Definition~\ref{def:metric-approx}).
\end{proof}

\begin{proposition} \label{prop:tight}
Let $k >\ell>1$. There are infinitely many values of $n$ for which there
exist HSTs (collectively denoted by $T$) with $n$ leaves such that:
\begin{enumerate}
\item Any $k$-HST that is $\ell$-approximated by a subspace of $T$, has at most
$n^{\frac{1}{\log_\ell k}}$ points.
\item Any ``binary/uniform" $k$-HST that is $\ell$-approximated by a
subspace of $T$, has at most $ 2^{2\sqrt{\frac{\log n}{\log_\ell k} }}$
points.
\item Any ``KRR" $k$-HST that is $\ell$-approximated by a
subspace of $T$, has at most $O(\frac{\log n}{\log_\ell k \log \log n})$
points.

\item Any ``BFM" HST that is $\ell$-approximated by a subspace of $T$,
has at most $O(\log n)$ points.
\end{enumerate}
\end{proposition}
\begin{proof}
The examples will all have the same basic structure. Fix a small
constant $\e>0$ to be determined later, and let
$k>\ell'=(1+\e)\ell$. Consider an $\ell'$-HST $T$ such that an
internal vertex $v$ with edge depth $i$ has diameter
$\Delta(v)={\ell'}^{-i}$.  Let $h\in \mathbb{N}$ be a parameter of
the size of $T$.
\begin{enumerate}
\item  In this case, $T$ is a complete binary tree
of height $h\lceil \log_{\ell'}k \rceil $ with $n=2^{h \lceil \log_{\ell'}k
\rceil}$ leaves. Let $R\subseteq S$ be a subset of the points that $\ell$
approximates a $k$-HST $W$. Let $T'$ be the subtree of $T$ that its leaves
are exactly the subset $R$. It follows from Lemma~\ref{lem:iso-HSTs} that
the edge distance in $T'$ between any two non-degenerate vertices $u$ and
$v$ is at least $\lceil \log_ {\ell'} k\rceil $. Hence, when coalescing pair
of edges with common degenerate vertex in $T'$, the resulting tree is a
binary tree of height at most $h$ with the same set of leaves, $R$, and so
 \(|R|\leq 2^{h} \leq   n^{\frac{1}{\lceil \log_{\ell'}\rceil k}}.\)
Choosing $\e>0$ small enough  implies that $|R| <  n^{\frac{1}{\log_\ell k}}
+1$.

\item In this case, $T$ is a complete tree
of height $h \lceil\log_{\ell'} k\rceil$ and the out-degree of
each internal vertex is $2^h$. Hence $n=2^{h^2 \lceil\log_{\ell'}
k\rceil}$, so $h=\sqrt{\frac{\log n}{\lceil\log_{\ell'}k\rceil}}$.
Let $R$ be a subset of points approximating a ``binary/uniform"
$k$-HST $W$, and let $T'$ be the subtree of $T$ whose set of
leaves is exactly $R$. By Lemma~\ref{lem:iso-HSTs}, $T'$ is also a
``binary/uniform" HST. As before on any vertical path in $T'$
there are only $h$ non-degenerate vertices. After removing
degenerate vertices from $T'$ (by coalescing pair of edges with
common degenerate vertex), it is easy to show by induction on the
levels, that a vertex in level $i$ in $T'$ can not have more than
$2^{h-i}2^h$ leaves, and therefore $T'$ has no more than $2^{2h}$
leaves. By choosing $\e>0$ small enough we conclude the claim.

\item In this case, $T$ is a complete tree
of height $h \lceil \log_{\ell'} k \rceil$ and the out-degree of
each internal vertex is $h$. Assume also that $h\geq \lceil
\log_{\ell'} k \rceil$. Hence $n=h^{h\lceil \log_{\ell'}k\rceil}$,
so
 \[ h=\Theta \Bigl ( \frac{\log n}{\lceil\log_{\ell'}k\rceil (\log \log n -
 \log \lceil\log_{\ell'} k\rceil)} \Bigr )
  = \Theta(\frac{2 \log n}{\lceil\log_{\ell'}k\rceil \,\log \log n}). \]
The last inequality follows since $\log\log n\geq 2 \log
\lceil\log_{\ell'}k\rceil$.

Let $R$ be a subset of points approximating a ``KRR" $k$-HST $W$, and let
$T'$ be the subtree of $T$ that its leaves are exactly $R$. By
Lemma~\ref{lem:iso-HSTs}, $T'$ is also a ``KRR" $k$-HST. Either $T'$ is a
uniform metric, and then the leaves are all children of one vertex (after
removing degenerate vertices), and therefore there at most $h$ such leaves.
Otherwise, $T'$ is $k$-super-increasing. Only $h$ vertices on any vertical
path in $T'$ are non-degenerate. and so $T'$ has at most $h+1$ leaves. Again,
the claim follows by taking $\e>0$ small enough.

\item  In this case, $T$ is a complete binary tree
of height $h$ with $n=2^h$ leaves.
Let $R$ be a subset of points approximating a ``BFM" HST $W$, and let
$T'$ be the subtree of $T$ that its leaves are exactly $R$.
$T'$ is a binary tree, since it is a sub-tree of a binary tree.

We want to
prove that $T'$ is a ``BFM" HST. Assume for the sake of contradiction that
$T'$ is not a ``BFM" HST.
This implies
the existence of four distinct leaves $a,b,c,d$,  with the following
``pairing" property: There exists a partition of $a,b,c,d$ into
two pairs  $\{a,b\}$ and   $\{c,d\}$, such that
$\lca_{T'}(a,c)=\lca_{T'}(a,d)=\lca_{T'}(b,c)=\lca_{T'}(b,d)$ (equals $u$), but
 $\lca_{T'}(a,b)\neq u$ and $\lca_{T'}(c,d)\neq u$.
However, in a ``BFM" HST $S$, for any subset of leaves $A$, there exists
$x \in A$ such that for any $x\notin \{y,z\}\subset A$,
$\lca_S(x,y)=\lca_S(x,z)$. So without loss of generality,
$\lca_W(a,b)=\lca_W(a,c)=\lca_W(a,d)$.
By Proposition~\ref{prop:lca-in-HST}, it implies that
 $\lca_{T'}(a,b)=\lca_{T'}(a,c)=\lca_{T'}(a,d)$. Therefore any partition
of $a,b,c,d$ to two pairs contradicts the ``pairing" property.

So $T'$ is a ``BFM" tree, and therefore has at most $h+1$ leaves.
\end{enumerate}
\renewcommand{\qed}{}
\end{proof}

\Section{Concluding Remarks} \label{sec:open}

As mentioned before, the lower bound on the competitive ratio of the MTS
problem in $n$-point metric spaces was improved in
\cite{BLMN-phenomena} to $\Omega(\log n / \log \log n)$.
It is an interesting challenge to achieve $\Omega(\log n)$ lower
bound for any metric space.
A plausible way to do it is
proving a lower bound for $k$-HSTs
\emph{with constant} $k$. This was done in the context of proving
upper bounds on the competitive ratio for the MTS problem in
\cite{FM00} using ``fine grained" combining technique.

Lemma~\ref{lem:adv-composed} is a tight lower bound for UMTS on uniform
metric when assuming the conjecture of $\Theta(\log n)$ competitive ratio
for MTS. An interesting problem is to find a matching upper bound. This
would improve the general upper bound for MTS by a factor of $\log \log n$.
A harder problem is to improve the upper bound for MTS to $o(\log^2 n)$.

For the $K$-server problem in arbitrary metric spaces, no sub-linear upper
bound on the randomized competitive ratio is known.

\subsection*{Acknowledgments} We wish to thank Noga Alon, Amos Fiat, Guy Kindler, Nati
Linial, Yuri Rabinovich, Mike Saks, Steve Seiden, and Amit Singer
for many discussions and suggestions. In particular, Nati helped
in simplifying and improving earlier versions of
Lemma~\ref{lem:adv-all-cr} and
Proposition~\ref{prop:binary/balanced}.

\bibliographystyle{alpha}
\bibliography{lbrmts}

\appendix

\Section{Some Probabilistic Calculations} \label{sec:calc}

\newcommand{\inve}{f}

In this section we present some probabilistic arguments needed in
the proof of Lemma~\ref{lem:adv-all-cr}.


\begin{lemma} \label{lem:tail-lb}
There exist constants $\lambda_2\geq1\geq \lambda_1>0$ such that for any
binomial random variable $X$ with $p\leq 0.5$ and mean $\mu\geq 4$ and any
$\delta\in[0,1]$ we have
\[ \Pr [X\leq (1-\delta) \mu] \geq
\lambda_1 e^{- \lambda_2 \delta^2 \mu} .\]
\end{lemma}

Lemma~\ref{lem:tail-lb} is easily realized for most of the range of
$p,\delta,\mu$ using the Poisson and Normal approximations of binomial
distribution (cf. \cite[Ch.~1]{Bol85}). Here we give an elementary proof.

Set $\inve(x)= (1-x)^{1/x}$. Clearly,  $\inve$ is increasing as $x$
decreases to $0$, and its limit is $e^{-1}$.

\begin{proposition} \label{prop:app1}
Let $X\sim \mathcal{B}(m,p)$ be a Binomial random variable,
$p+q=1$, $p\leq 1/2$, $\mu=p m$, $k=(1-\eta)\mu \in [m]$, and
$\eta\in (0,1)$. Then
\[ \Pr[X=k] \geq \frac{\inve(\eta)^{\eta^2 \mu}}{3\sqrt{\mu}}.\]
\end{proposition}
\begin{proof}
Recall that by Stirling Formula (cf. \cite[pp. 4]{Bol85}),

\begin{align*}
\Pr[X=k] &= \binom{m}{k} p^k q^{m-k}
  \geq \frac{1}{e^{1/6}\sqrt{2\pi k}} \bigl( \frac{p m}{k} \bigr )^k
  \bigl( \frac{q m}{m-k} \bigr )^{(m-k)}
   \geq \frac{1}{3\sqrt{\mu}} \bigl( \frac{p m}{k} \bigr )^k
  \bigl( \frac{q m}{m-k} \bigr )^{(m-k)}
\end{align*}
Also,

\begin{align*}
\bigl( \frac{p m}{k} \bigr )^k &= \bigl( \frac{\mu}{(1-\eta) \mu} \bigr
)^{(1-\eta)\mu}
     = (1-\eta)^{- (1-\eta)\mu} = f(\eta)^{-\eta (1-\eta) \mu}
     =f(\eta)^{-\eta\mu}\, f(\eta)^{\eta^2 \mu}
\end{align*}

\begin{align*}
\bigl( \frac{q m}{m-k} \bigr )^{(m-k)}
= \bigl( 1 -\frac{\eta p}{q +\eta p} \bigr )^{m (q +\eta p)} & =
f(\frac{\eta p}{q +\eta p}) ^ {\frac{\eta p}{q +\eta p} \,  m (q +\eta p)} =
f(\frac{\eta p}{q +\eta p}) ^{\eta \mu}
\end{align*}

Note that $\eta \geq (\eta p)/(q +\eta p)$, so $f(\eta) \leq
f((\eta p)/(q +\eta p))$, and the claim is proved.
\end{proof}

\begin{proposition} \label{prop:taillb}
Given a binomial random variable $X$ with mean $\mu$, $\delta \leq
1/3$, and $\delta \mu \geq 4$, then
\[ \Pr [X\leq (1-\delta) \mu] \geq
  \frac{\delta \sqrt{\mu}}{6}\; e^{- 7 \delta^2 \mu} .\]
\end{proposition}
\begin{proof} Applying Proposition~\ref{prop:app1},
\begin{align*}
\Pr [X\leq (1-\delta) \mu] &\geq \sum_{k=\lceil(1-2\delta)\mu
\rceil}^{\lfloor(1-\delta)\mu \rfloor} \Pr[X=k] \geq
(\delta \mu -2)\Pr[X = \lceil(1-2\delta) \mu \rceil] \\
 & \geq  \frac{\delta \mu}{2}  \,
       \frac{(\inve(2\delta))^{2^2\delta^2 \mu}}{ 3\sqrt{\mu}}
  \geq \frac{\delta \sqrt{\mu}}{6}\; {3}^{- 1.5\cdot 4  \delta^2 \mu}
  \geq \frac{\delta \sqrt{\mu}}{6}\; e^{- 7 \delta^2 \mu} . \qquad \qed
\end{align*}
\renewcommand{\qed}{}
\end{proof}

\begin{proof}[Proof of Lemma~\ref{lem:tail-lb}]
For $\delta>1/3$:
 \[ \Pr[X\leq (1-\delta)\mu ]\geq \Pr [X=0 ] \geq
  (1-p)^m = ((1-p)^{p^{-1}})^\mu  \geq  4^{-
\mu} \geq e^{- 13\delta^2 \mu} . \]
 For $4 \leq \mu\leq 12^2$ and $\delta
\leq 1/3$: There exists $\delta'\in[\delta,\delta+\mu^{-1})$  such
that $(1-\delta')\mu \in \mathbb{N}$,  so $\delta'\leq
\delta+\tfrac{1}{4} \leq \tfrac{2}{3}$. Applying
Proposition~\ref{prop:app1},
 \[ \Pr [X\leq (1-\delta)\mu ] \geq \Pr [X = (1-\delta')\mu ] \geq
 \frac {(1/3)^{1.5 \delta'{}^2 \mu}}{36}\geq \frac{e^{-1.7
 (\delta^2 \mu+ 2\delta + \mu^{-1})}}{36} \geq \frac{e^{-1.7\delta^2 \mu}}{5\cdot 36} . \]
For $\mu\geq 12^2$ and $\tfrac{1}{3}\geq \delta \geq
\tfrac{1}{3}\mu^{-0.5}$: we note that $\delta \mu \geq 4$, so
applying Proposition~\ref{prop:taillb},
 \[ \Pr[X\leq (1-\delta)\mu] \geq \frac{1}{18}\; e^{- 7 \delta^2 \mu} .\]
For $\mu \geq 12^2$ and $ \tfrac{1}{3}\mu^{-0.5} \geq \delta$: let
$\delta'= \tfrac{1}{3}\mu^{-0.5}$. Note that $\tfrac{1}{3}\geq
\delta' \geq \delta$,  so applying Proposition~\ref{prop:taillb},
 \[ \Pr[ X\leq (1-\delta)\mu ] \geq \Pr[ X\leq (1-\delta')\mu ] \geq
  \frac{e^{-\tfrac{7}{9}}}{18} \geq  \frac{1}{40}  .\]

We conclude that $\Pr[ X\leq (1-\delta)\mu ] \geq
\tfrac{1}{180}e^{-13\delta^2 \mu}$.
\end{proof}

\begin{proposition} \label{prop:depen}
Consider the following experiment: $m$ balls are randomly put in $n$ bins.
Let $X_i$ be the number of balls in the $i$th bin. Then, for any $1\leq \ell
\leq n$ and any integer sequence $(\alpha_i)_{1\leq i \leq \ell}$,
\[ \Pr [ \bigwedge _{i=1}^\ell (X_i > \alpha_i)] \leq \prod_{i=1}^ \ell \Pr[X_i
> \alpha _i] .\]
\end{proposition}
\begin{proof}
Let $E_i$ be the event $X_i >\alpha_i$. Fixing $i>1$, let $a_j=\Pr [E_1\land
\cdots \land E_{i-1}\; | \; X_i =j ]$. It is elementary to check that $a_j$
is monotonic non-increasing in $j$. Thus,
\begin{align*}
\Pr [E_1\land \cdots \land E_{i-1}\; | \; E_i]
 & =\sum_{j>\alpha_i} a_j \frac{\Pr[X_i=j]}{\Pr[E_i]}
  \leq \sum_{j\leq \alpha_i} a_j \frac{\Pr[X_i=j]}{1-\Pr[E_i]},
\end{align*}
and so
\begin{align}
\Pr [E_1\land \cdots \land E_{i-1}\; | \; E_i] & \leq
  \Pr[E_i] \sum_{j>\alpha_i} a_j \frac{\Pr[X_i=j]}{\Pr[E_i]}
 +(1-\Pr[E_i]) \sum_{j\leq \alpha_i} a_j \frac{\Pr[X_i=j]}{1-\Pr[E_i]} \nonumber\\
 & = \sum_j a_j \Pr[X_i=j]
 = \Pr[E_1 \land \cdots \land E_{i-1}  ]. \label{eq:depen}
\end{align}

We conclude by induction on $i$ that $\Pr[E_1 \land \cdots \land E_n] \leq
\Pr[E_1] \Pr[E_2] \cdots \Pr[E_i]$, since
\begin{align*}
\Pr[E_1 \land \cdots \land E_i] &= \Pr[E_1 \land \cdots \land E_{i-1} |
E_i]\; \Pr[E_{i}] \\
&\leq \Pr[E_1 \land \cdots \land E_{i-1}] \Pr[E_{i}]  \leq \Pr[E_1] \Pr[E_2]
\cdots \Pr[E_i]
\end{align*}
The last inequality follows from the induction hypothesis.
\end{proof}

\begin{proposition} \label{prop:depen2}
Under the conditions of Proposition~\ref{prop:depen}, given $\alpha>0$,
denote by $Z = \bigwedge_{i=2}^n (X_i>\alpha)$, then
 \( E[X_1|Z] \leq E[X_1].  \)
\end{proposition}
\begin{proof}
 From Eq.~\eqref{eq:depen}  \emph{in the proof of}
Proposition~\ref{prop:depen},
 \[ \Pr[X_1>j| Z]= \frac{\Pr [(X_1>j) \wedge \bigwedge_{i=2}^n
 (X_i>\alpha)]}{\Pr [\bigwedge_{i=2}^n (X_i>\alpha)]}\leq \Pr[X_1>j] .\]
In general, for integer non negative variable $W$, we have that
 \[ E[W]= \sum _{j=0}^\infty j \Pr[W\!=\!j]=  \sum _{j=0}^\infty j \bigl(\Pr[W>j-1]-
 \Pr[W>j] \bigr )= \sum_{j=0}^\infty \Pr[W>j] ,\]
so in our case,
\[ E[X_1|Z]= \sum_{j=0}^\infty \Pr[X_1>j| Z]\leq \sum_{j=0}^\infty \Pr[X_1>j]= E[X_1].
\qquad \qquad \qed \]
\renewcommand{\qed}{}
\end{proof}

\end{document}